\documentclass[aps,pre,reprint,showpacs,floatfix,groupedaddress]{revtex4-1}

\usepackage{amscd}
\usepackage{amsmath}
\usepackage{amssymb}
\usepackage{amsthm}
\usepackage{bm}
\usepackage{epsfig}
\usepackage{epstopdf}
\usepackage{graphicx}
\usepackage{latexsym}
\usepackage{mathtools}
\usepackage{mathrsfs}
\usepackage{subfigure}
\usepackage{appendix}
\usepackage{multirow}

\makeatletter
\DeclareRobustCommand*\textsubscript[1]{%
  \@textsubscript{\selectfont#1}}
\def\@textsubscript#1{%
  {\m@th\ensuremath{_{\mbox{\fontsize\sf@size\z@#1}}}}}
\makeatother

\begin{document}

\title{Selectively Localized Wannier Functions}
\author{Runzhi Wang$^1$, Emanuel A. Lazar$^2$, Hyowon Park$^{1,3}$, Andrew J. Millis$^1$, Chris A. Marianetti$^3$ }
\affiliation
{
 $^1$Department of Physics, Columbia University, New York, New York 10027.\\
 $^2$Materials Science and Engineering, University of Pennsylvania, Philadelphia, PA 19104.\\
 $^3$Applied Physics and Applied Mathematics, Columbia University, New York, New York 10027.
} 
\date{\today}

\begin{abstract}
Since the seminal work of Marzari and Vanderbilt, maximally localized Wannier functions  have become widely used as a real-space representation of the electronic structure of periodic materials.  In this paper we introduce selectively localized Wannier functions  which extend the method of Marzari and Vanderbilt in two important ways.  First, our method allows us to focus on localizing a subset of orbitals of interest. Second, our method allows us to fix centers of these orbitals, and ensure the preservation of the point-group symmetry. These characteristics are important when  Wannier functions are used in methodologies that go beyond density functional theory by treating a local subspace of the Hamiltonian more effectively. Application of our method to GaAs, SrMnO$_3$, and Co demonstrates that selectively localized Wannier functions  can offer improvements over the maximally localized Wannier function technique. 
\end{abstract}


\maketitle

\section{Introduction}
Solving the Schrodinger equation in crystalline solids typically relies on the use of the  translation group to block-diagonalize the Hamiltonian, yielding Bloch states $\psi_{n\bf k}$, where $\bf k$ is a wave vector in the first Brillouin zone of the solid  and $n \in \{1, 2 \ldots  J\}$ is a band index. However, it is often both physically and computationally desirable to have a real space, atomic-like representation of the Hamiltonian. An approach to achieve this was introduced by Wannier in 1937 \cite{wannier1937structure}.  
Considering a single band, Wannier introduced the following transformation:
\begin{equation}
|{\bf R} n \rangle = \frac{V}{(2\pi)^3}\int d{\bf k} e^{-i{\bf k} \cdot {\bf R}} |\psi_{n \bf k}\rangle,
\label{eq1}
\end{equation}
so that
\begin{eqnarray}
|\psi_{n \bf k}\rangle = \sum_{\bf R} e^{i{\bf k} \cdot {\bf R}}|{\bf R} n \rangle.
\end{eqnarray}
where $\bf R$ is any lattice vector and $V$ is the unit cell volume. The function $\langle {\bf r} |{\bf R} n \rangle$ is referred to as a Wannier function. 

Because each Bloch function $\psi_{n\bf k}$ can be adjusted by an arbitrary phase $e^{i \varphi_{n{\bf k}}}$, the Wannier functions are not uniquely defined, and we can rewrite Equation \ref{eq1} as:
\begin{equation}
|{\bf R} n \rangle = \frac{V}{(2\pi)^3}\int d{\bf k} e^{-i{\bf k} \cdot {\bf R}} e^{i\varphi_{n{\bf k}}} |\psi_{n \bf k}\rangle  
\end{equation}
If Wannier functions are constructed from a set of $J$ bands, there is an even greater freedom, since at each $k$-point we may use an arbitrary unitary transformation of band states.  Denoting an arbitrary $J \times J$ unitary matrix as $U^{\bf k}$, then at a given $k$-point we are free to mix the Bloch states as follows:
\begin{equation}
|\tilde \psi_{n\bf k} \rangle = \sum_{m=1}^J U^{\bf k}_{mn} |\psi_{m\bf k}\rangle
\end{equation}
by which we can define generalized Wannier functions:
\begin{equation}
|{\bf R} n \rangle = \frac{V}{(2\pi)^3}\int d{\bf k} e^{-i{\bf k} \cdot {\bf R}} |\tilde \psi_{n \bf k}\rangle
\end{equation}
This ``gauge freedom'' may be exploited to define Wannier functions that are optimized for particular purposes.  One approach, pioneered by Marzari and Vanderbilt \cite{marzari1997maximally}, is to choose the  $U^{\bf k}$  such that the complete set of Wannier functions are as localized as possible in the position representation.  Specifically, if we let
\begin{equation}
\bar{\bf r}_n = \langle {\bf 0}n | {\bf r} | {\bf 0}n \rangle
\end{equation}
and
\begin{equation}
\langle r^2 \rangle_n = \langle {\bf 0}n | r^2 | {\bf 0}n \rangle
\end{equation}
then we can define the total spread functional $\Omega$ of the $J$ Wannier functions $|{\bf R} n \rangle$ as:
\begin{equation}
\Omega = \sum_{n=1}^J \left[ \langle r^2 \rangle_n - \bar{\bf r}_n^2 \right]
\label{Omegadef}
\end{equation}
and choose the matrices $U^{\bf k}$ by minimizing  $\Omega$ \cite{marzari1997maximally}.  The functions resulting from this procedure are termed  maximally localized Wannier functions (MLWFs). In addition to introducing the MLWF method, Marzari and Vanderbilt introduced a gradient descent method for performing the minimization.

While MLWFs are now very widely used, the global spread function given in Eq.~\ref{Omegadef} is not necessarily optimal for all uses. In recent years attention has focussed on methodologies that go beyond density functional theory (DFT) by more appropriately treating electronic correlations in some relevant local subspace of the Hamiltonian. For example, in the  DFT+U and DFT plus dynamical mean-field theory (DMFT) methods one treats beyond DFT correlations in a local subspace corresponding to atomic-like $d$ (for transition metals or transition metal oxides) or $f$ (for rare earth or actinide intermetallics) orbitals.  In these applications, it is important to have a correct local description of the local subspace, including both the correct location of the centers of the correlated orbitals and the correct point symmetry,  but the properties of the other degrees of freedom are irrelevant. The MLWF procedure, which treats all orbitals on an equal footing, may give a sub-optimal description of the orbitals of interest and in particular does not guarantee that the centers and point symmetries of the orbitals of interest are correctly described. It is important to note that the method of Wannier functions for entangled energy bands \cite{souza2001maximally} also does not fulfill these goals. In this paper we develop a technique which allows the selective localization of  a subset of Wannier orbitals, with specified  Wannier center and point symmetry. 

The remainder of this paper is outlined as follows. In section II, we outline our methodology. In section III, we illustrate the utility of our method in the one-dimensional chain. In section IV, we present applications to GaAs, SrMnO$_3$, and Co. In section V, we analyze the Hamiltonian in the Wannier basis for SrMnO$_3$ and Co.

\section{Method}
Here we derive all of the relevant equations for our formalism. For simplicity we present a  construction in which the total number of Wannier functions is equal to the total number, J,  of bands under consideration.  We believe that our method can be extended to incorporate the inner and outer window construction of Souza et al [3] but this extension is not attempted in this paper because it does not appear to be necessary for the applications we  consider.  We formulate the problem using a ``band window'' construction which in principle requires no assumptions about separation of bands.  However, the applications we envisage (for example to DMFT calculations)  require a basis set that faithfully represents the charge density. Therefore, it is essential that the lowest included band is
separated from lower bands by an energy gap, and in what follows we choose windows such that this is the case. 

In the rest of this section we first describe the selective
localization of a subset of $J^\prime<J$ orbitals, then we present the
procedure for fixing the centers of the localized orbitals and finally explain
how we constrain the symmetry of the localized orbitals. In the remainder of
the paper we use MLWF to refer to the maximal localization method
\cite{marzari1997maximally, marzari2012maximally}; SLWF to refer to the
selectively localized Wannier method presented here; SLWF+C refers to the same
method with centers fixed; SLWF+CS refers to the same method with both centers
and symmetries fixed, as described in the Appendix.  Our SLWF method produces
two types of Wannier functions: objective Wannier functions (OWFs)  which have
a minimum cumulative spread and the remaining Wannier functions to which we assign no
specific name.

\subsection{Selective localization}
We construct a subset of localized orbitals by minimizing the following functional
\begin{equation}
\Omega = \sum_{n=1}^{J'} [\langle r^2 \rangle_n - \bar{\bf r}^2_n]
\label{omega}
\end{equation}
where $J'\leq J$ is the number of objective Wannier functions that we choose; recall that $J$ is the total number of Wannier functions, or equivalently the number of bands considered. Our method reduces to MLWF when $J'=J$.

Marzari and Vanderbilt showed \citep{marzari1997maximally} that in the case $J^\prime=J$, $\Omega$ can be decomposed into the sum of two terms, one of which is invariant under arbitrary unitary transformations.  However, when $J' < J$, this is no longer the case, but the minimization of the functional can still be accomplished by methods very similar to those   of Marzari and Vanderbilt. 

We write $\Omega = \Omega_{IOD}+\Omega_D$, where
\begin{eqnarray}
\Omega_{IOD} &=& \sum_{n=1}^{J'}[ \langle r^2\rangle_n- \sum_{\mathbf R}|\langle{{\mathbf R}n}|{\mathbf r}|{{\mathbf 0}n}\rangle|^2]\\
\Omega_D &=& \sum_{n=1}^{J'}\sum_{{\mathbf R}\neq {\mathbf 0}}|\langle{{\mathbf R}n}|{\mathbf r}|{{\mathbf 0}n}\rangle|^2
\end{eqnarray} 
Following Ref.~\citep{marzari1997maximally} we recast the expression as a discretized sum in $k$-space:
\begin{eqnarray}
\Omega_{IOD}&=&\frac{1}{N}\sum_{n=1}^{J'}\sum_{\mathbf k, \mathbf b}w_b(1-|M_{nn}^{\mathbf k, \mathbf b}|^2)\\
\Omega_D &=& \frac{1}{N}\sum_{n=1}^{J'}\sum_{\mathbf k, \mathbf b}w_b( \operatorname{Im}\ln M_{nn}^{\mathbf k, \mathbf b} + \mathbf b \cdot \bar{\mathbf r}_n)^2
\end{eqnarray}
where $M_{mn}^{\mathbf k, \mathbf b} \equiv \langle u_{m \bf k}|u_{n {\bf k}+{\bf b}}\rangle$, $\bf b$ are vectors which connect a $k$-point to its near neighbors, $w_b$ is a weight for each $|{\bf b}| = b$ such that $\sum_{\bf b} w_b b_\alpha b_\beta = \delta_{\alpha \beta}$  (see Appendix B of Ref.~\citep{marzari1997maximally} for a detailed explanation).

Under the infinitesimal unitary transformation, $U_{mn}^{\mathbf{k}}=\delta_{mn}+dW_{mn}^{\mathbf{k}}$, where $dW^{\mathbf{k}\dagger}=-dW^{\mathbf{k}}$, the wave functions transform as
\begin{eqnarray}
|u_{n\mathbf k}\rangle &\rightarrow& |u_{n\mathbf k}\rangle+\sum_{m=1}^J dW_{mn}^{\mathbf{k}} |u_{m\mathbf k}\rangle
\end{eqnarray}
so that 
\begin{eqnarray}
d\Omega_{IOD} &=& \frac{4}{N}\sum_{\mathbf k, \mathbf b}w_b\sum_{n=1}^{J'}\sum_{m=1}^J \operatorname{Re}(dW_{nm}^{\mathbf k}R_{mn}^{\mathbf k, \mathbf b})\\
d\Omega_D &=& -\frac{4}{N}\sum_{\mathbf{k},\mathbf{b}}w_b\sum_{n=1}^{J'}\sum_{m=1}^{J}\operatorname{Im}(dW_{nm}^{\mathbf{k}}T_{mn}^{\mathbf k, \mathbf b})
\end{eqnarray}
where 
\begin{eqnarray}
R_{mn}^{\mathbf{k},\mathbf{b}} &=& M_{mn}^{\mathbf{k},\mathbf{b}} M_{nn}^{\mathbf{k},\mathbf{b}*}
\end{eqnarray}
\begin{eqnarray}
T_{mn}^{\mathbf{k},\mathbf{b}} &=& \tilde{R}_{mn}^{\mathbf{k},\mathbf{b}} (\operatorname{Im}\ln M_{nn}^{\mathbf k, \mathbf b} + \mathbf b \cdot \bar{\mathbf r}_n )
\end{eqnarray}
\begin{eqnarray}
\tilde{R}_{mn}^{\mathbf{k},\mathbf{b}} &=& M_{mn}^{\mathbf{k},\mathbf{b}}/M_{nn}^{\mathbf{k},\mathbf{b}}
\end{eqnarray}
Thus, the gradient of the spread functional is:
\begin{eqnarray}
G_{mn}^{\mathbf{k}}&=&\frac{d\Omega}{dW_{nm}^{\mathbf{k}}}\\
&=&\begin{dcases}
4\sum_{\mathbf b}w_b(\mathscr{A}[R_{mn}^{\mathbf k, \mathbf b}]-\mathscr{S}[T_{mn}^{\mathbf k, \mathbf b}]) &m\leq J', n\leq J'\\
-4\sum_{\mathbf b}w_b\left( \frac{R^{\mathbf k, \mathbf b*}_{nm}}{2}+\frac{T^{\mathbf k, \mathbf b*}_{nm}}{2i}\right)&  m\leq J', n>J'\\
4\sum_{\mathbf b}w_b\left(\frac{R_{mn}^{\mathbf k, \mathbf b}}{2}-\frac{T_{mn}^{\mathbf k, \mathbf b}}{2i}\right)& m>J', n\leq J'\\
0 & m>J', n>J'\nonumber
\end{dcases}
\end{eqnarray}
where $\mathscr{A}[R_{mn}^{\mathbf k, \mathbf b}]= (R_{mn}^{\mathbf k, \mathbf b}-R_{nm}^{\mathbf k, \mathbf b*})/2$, $\mathscr{S}[T_{mn}^{\mathbf k, \mathbf b}]= (T_{mn}^{\mathbf k, \mathbf b}+T_{nm}^{\mathbf k, \mathbf b*})/2i$.

Following Ref.~\citep{marzari1997maximally}, we minimize $\Omega$ by updating $U^{\mathbf{k}}$ in small steps according to $\exp\left[dW^{\mathbf{k}}\right]$, choosing:
\begin{eqnarray}
dW^{\mathbf{k}}=\epsilon G^{\mathbf{k}}
\end{eqnarray}
where $\epsilon$ is a positive infinitesimal. We thus have:
\begin{eqnarray}
d\Omega&=&\sum_{\mathbf{k}}\sum_{m,n=1}^J G_{mn}^{\mathbf{k}}dW_{nm}^{\mathbf{k}}\nonumber\\
&=&-\epsilon\sum_{\mathbf{k}}\sum_{m,n=1}^J |G_{mn}^{\mathbf{k}}|^2,
\end{eqnarray}
using the identity $G^\dagger = -G$. Thus, it is guaranteed that $d\Omega \leq 0$. This allows us to iteratively update the unitary matrix until a converged solution is attained. In practice, we fix the step size by choosing $\epsilon=\alpha / 4w$, where $w=\sum_{\mathbf{b}}w_{b}$, and minimize the spread using a nonlinear conjugate gradient method \cite{klockner2004computation}.  While this method finds only local minima, we have found in practice that if a reasonable starting point is chosen, physically reasonable minima are found and we believe these are global minima based on substantial testing.

\subsection{Fixing centers}
To fix the centers of our objectively localized Wannier functions we introduce a Lagrange multiplier to constrain the Wannier center $\bar{\mathbf{r}}_n$:
\begin{eqnarray}
\lambda_c \sum^{J'}_{n=1}(\bar{\mathbf{r}}_n-\mathbf{r}_{0n})^2
\end{eqnarray}
where $\mathbf{r}_{0n}$ is the desired center for the $n$th Wannier function, and $\lambda_c$ is a Lagrange multiplier for this constraint.  Here $J'$ is chosen to allow for a selective localization.  We impose this constraint into $\Omega$ which is defined by Equation \ref{omega} and introduce a new target functional:
\begin{eqnarray}
\label{center_objective}
\Omega_c &=& \sum^{J'}_{n=1}\left[\langle  r^2 \rangle_n - \bar{\mathbf r} ^2_n+\lambda_c(\bar{\mathbf{r}}_n-\mathbf{r}_{0n})^2\right]
\end{eqnarray}

We  decompose $\Omega_c$ in a manner similar to that in the previous subsection, but with an additional term $\Omega_{c,\nu}$ that results from the imposed constraint:
\begin{eqnarray}
\Omega_c&=&\Omega_{c,IOD}+\Omega_{c,D}+\Omega_{c,\nu}
\end{eqnarray}
where:
\begin{eqnarray}
\Omega_{c,IOD}&=& \sum_{n=1}^{J'}[\langle r^2\rangle_n- (1-\lambda_c)\sum_{\mathbf R}|\langle{{\mathbf R}n}|{\mathbf r}|{{\mathbf 0}n}\rangle|^2] 
\end{eqnarray}
\begin{eqnarray}
\Omega_{c,D}&=& (1-\lambda_c)\sum_{n=1}^{J'}\sum_{{\mathbf R}\neq {\mathbf 0}}|\langle{{\mathbf R}n}|{\mathbf r}|{{\mathbf 0}n}\rangle|^2 \\
\Omega_{c,\nu}&=&\lambda_c \sum_{n=1}^{J'} \mathbf{r}_{0n}^2-2\lambda_c\sum_{n=1}^{J'}\mathbf{r}_{0n}\cdot\bar{\mathbf{r}}_n
\end{eqnarray}
We can recast the expression as a discretized sum in $k$-space:
\begin{eqnarray}
\Omega_{c,IOD} &=& \frac{1}{N}\sum_{n=1}^{J'}\sum_{\mathbf k, \mathbf b}w_b\left[1-|M_{nn}^{\mathbf k, \mathbf b}|^2+\lambda_c(\operatorname{Im}\ln M_{nn}^{\mathbf{k},\mathbf{b}})^2\right] \nonumber \\
 &\label{whatever2}\\
\Omega_{c,D} &=& (1-\lambda_c)\frac{1}{N}\sum_{n=1}^{J'}\sum_{\mathbf k, \mathbf b}w_b( \operatorname{Im}\ln M_{nn}^{\mathbf k, \mathbf b} + \mathbf b \cdot \bar{\mathbf r}_n)^2 \nonumber \\
 &\label{whatever3}\\
\Omega_{c,\nu} &=& \lambda_c \sum_{n=1}^{J'} \mathbf{r}_{0n}^2+\lambda_c\frac{2}{N}\sum_{\mathbf k, \mathbf b} w_b \mathbf b\cdot\sum_{n=1}^{J'} \mathbf{r}_{0n}\operatorname{Im}\ln M_{nn}^{\mathbf k, \mathbf b} \nonumber \\
 &\label{whatever4}
\end{eqnarray}
Under the infinitesimal unitary transformation, we have:
\begin{eqnarray}
d\Omega_{c,IOD}&=& \frac{4}{N}\sum_{\mathbf k, \mathbf b}w_b\sum_{n=1}^{J'}\sum_{m=1}^J [\operatorname{Re}(dW_{nm}^{\mathbf k}R_{mn}^{\mathbf k, \mathbf b})\nonumber\\
&&-\lambda_c\operatorname{Im}\ln M_{nn}^{\mathbf k, \mathbf b}\operatorname{Im} ( dW_{nm}^{\mathbf{k}} \tilde{R}_{mn}^{\mathbf{k},\mathbf{b}} ) ]\\
d\Omega_{c,D}&=&-(1-\lambda_c)\frac{4}{N}\sum_{\mathbf{k},\mathbf{b}}w_b\sum_{n=1}^{J'}\sum_{m=1}^{J}\operatorname{Im}(dW_{nm}^{\mathbf{k}}T_{mn}^{\mathbf k, \mathbf b}) \nonumber \\
 &\label{whatever5}\\
d\Omega_{c,\nu}&=&-\lambda_c\frac{4}{N}\sum_{\mathbf k, \mathbf b} w_b \mathbf b\cdot\sum_{n=1}^{J'}\sum_{m=1}^J \mathbf{r}_{0n}\operatorname{Im} (dW_{nm}^{\mathbf{k}} \tilde{R}_{mn}^{\mathbf{k},\mathbf{b}}) \nonumber \\
 &\label{whatever6}
\end{eqnarray}
Thus, the gradient of the functional is:
\begin{widetext}
\begin{eqnarray}
G_{c,mn}^{\mathbf{k}}&=&\frac{d\Omega_c}{dW_{nm}^{\mathbf{k}}}\\
&=&\begin{dcases}
4\sum_{\mathbf b}w_b\left\lbrace \mathscr{A}[R_{mn}^{\mathbf k, \mathbf b}]-(1-\lambda_c)\mathscr{S}[T_{mn}^{\mathbf k, \mathbf b}]\right\rbrace \\ \hspace{5mm}-4\lambda_c\sum_{\mathbf b}w_b\left[\frac{\tilde R_{mn}^{\mathbf k, \mathbf b}}{2i}\operatorname{Im}\ln M_{nn}^{\mathbf{k},\mathbf{b}}\right. \left.+\frac{\tilde R_{nm}^{\mathbf k, \mathbf b*}}{2i}\operatorname{Im}\ln M_{mm}^{\mathbf{k},\mathbf{b}}+\mathbf{b}\cdot \left(\mathbf{r}_{0n} \frac{\tilde{R}_{mn}^{\mathbf k, \mathbf b}}{2i}+ \mathbf{r}_{0m} \frac{\tilde{R}_{nm}^{\mathbf k, \mathbf b*}}{2i}\right)\right] & m\leq J', n\leq J'\\
-4\sum_{\mathbf b}w_b\left[\frac{R^{\mathbf k, \mathbf b*}_{nm}}{2}+(1-\lambda_c)\frac{T^{\mathbf k, \mathbf b*}_{nm}}{2i}\right] -4\lambda_c\sum_{\mathbf b}w_b \left(\frac{\tilde R_{nm}^{\mathbf k, \mathbf b*}}{2i}\operatorname{Im}\ln M_{mm}^{\mathbf{k},\mathbf{b}}+\mathbf{b}\cdot\mathbf{r}_{0m} \frac{\tilde R_{nm}^{\mathbf k, \mathbf b*}}{2i} \right)& m\leq J', n>J'\\
4\sum_{\mathbf b}w_b\left[\frac{R_{mn}^{\mathbf k, \mathbf b}}{2}-(1-\lambda_c)\frac{T_{mn}^{\mathbf k, \mathbf b}}{2i}\right]-4\lambda_c\sum_{\mathbf b}w_b\left(\frac{\tilde R_{mn}^{\mathbf k, \mathbf b}}{2i}\operatorname{Im}\ln M_{nn}^{\mathbf{k},\mathbf{b}}+\mathbf{b}\cdot \mathbf{r}_{0n} \frac{\tilde R_{mn}^{\mathbf k, \mathbf b}}{2i}\right) & m>J', n\leq J'\\
0& m>J',n>J'\nonumber\\
\end{dcases}
\end{eqnarray}
\end{widetext}
Minimizing this modified functional, we obtain Wannier functions that are maximally localized subject to the constraint of fixed centers.  Although this constraint is satisfied at the cost of some delocalization, we can still maintain a high degree of localization through concurrent selective localization, as we illustrate in the applications below. 

\subsection{Fixing Symmetry}
It is further possible to ensure that the Wannier functions obtained preserve not only the desired centers, but also that they  transform as irreducible representations of the point group, using for example the elegant group-theory based approach  recently introduced by Sakuma \cite{sakuma2013symmetry}. More straightforwardly one may simply introduce additional Lagrange multipliers, as discussed in the Appendix for the case of a one-dimensional model system.  However, we have found that in all of the realistic examples studied in this paper orbitals which are localized to a site with a given point group symmetry automatically transform as an  irreducible representation  of the respective point group. We do not presently have an analytical understanding of this empirical observation.

\section{One-dimensional chain}
\label{positive-periodic}
To illustrate the utility of our method, we  consider in this section a one-dimensional periodic lattice with $\delta$ potential barriers. In this system, there are an infinite number of isolated bands. As an example, we choose to construct a manifold of $4$ Wannier functions (ie. $4$ bands) with $2$ objective Wannier functions. We use a mesh with 100 $k$-points for MLWF, SLWF, and SLWF+C.  While considering examples in which symmetries are enforced (SLWF+CS), we use a mesh with 20 $k$-points, due to the increased computational demands involved in those cases.  In this one-dimensional problem, we study the following
Hamiltonian:
\begin{equation}
H=-\frac{a^2}{2}\left(\frac{d}{dx}\right)^2+\beta a\sum_j\delta(x-ja)
\label{H1d}
\end{equation}
where $a$ is the lattice constant, $\beta$ is the dimensionless ``strength" for the $\delta$-function.
In practice, we choose $a=5\AA$, $\beta=0.6610$.  

\begin{figure}
\begin{center}
\includegraphics[width=\linewidth,clip= ]{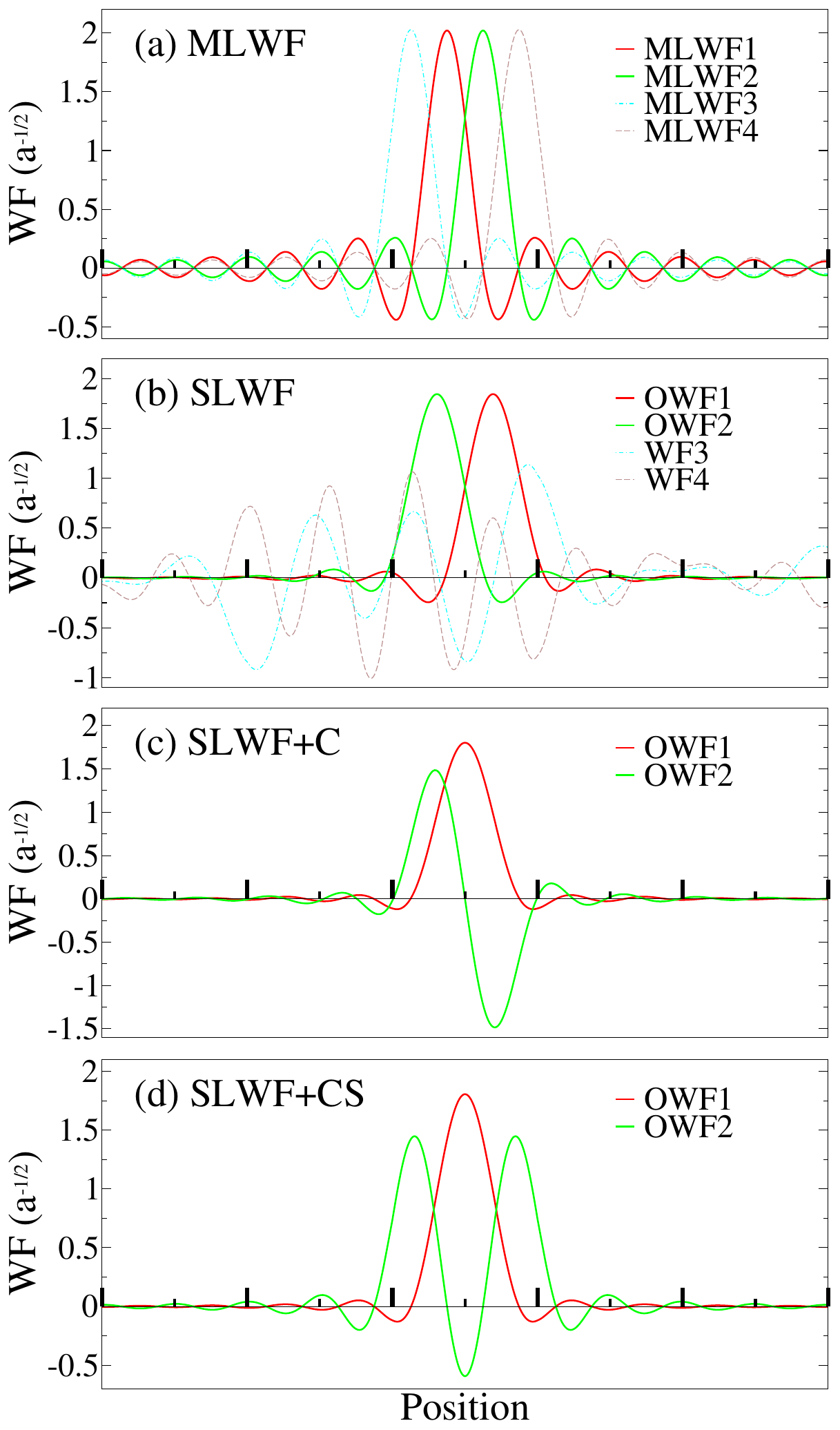}
\end{center}
\caption{Wannier functions for the 1-d chain of $\delta$-function potentials. Large tick marks denote the $\delta$-function, while small tick marks denote the midpoint.
({\bf a}) MLWF for $J = J'=4$. $\Omega_1 = \Omega_4 = 0.2853a^2, \Omega_2 = \Omega_3 = 0.2983a^2$.
({\bf b}) SLWF for $J = 4, J' = 2$. $\Omega_1=0.0162a^2$,  $\Omega_2=0.0162a^2$.
({\bf c}) SLWF+C for $J = 4, J' = 2$. Both are located at $x=0.5a$; $\Omega_1=0.0154a^2$,  $\Omega_2=0.0605a^2$.
({\bf d}) SLWF+CS for $J=4, J'=2$.  Both are symmetric about $0.5a$; $\Omega_1=0.0154a^2, \Omega_2=0.1387a^2$. 
}
\label{wannier_1d}
\end{figure}

Figure \ref{wannier_1d}(a) shows the Wannier functions resulting from MLWF (i.e. $J^\prime=J=4$). Three potential drawbacks are apparent.  First, the MLWFs are nearly equally localized, which results in each MLWF having a relatively long tail.  Second, we observe that none of the MLWFs are centered at either $x=0$, the location of the periodic $\delta$-potential, or $x=0.5a$, the midpoint between adjacent $\delta$-functions. Third, we notice that the MLWFs do not transform as irreducible representations of the order $2$ point group at $x=0$ nor $x=0.5a$. 

Panel (b) of Figure \ref{wannier_1d} shows the results of selective localization of two orbitals. Compared with MLWFs, the OWFs (solid lines, red and green on line)  become noticeably more localized (compare the tails of the functions in panels (a) and (b)).  The numerically computed spread $\langle (r-\langle r\rangle)^2\rangle$ of the most localized MLWF is 0.2853$a^2$ while the OWF has a spread of 0.0162$a^2$.  However, the remaining Wannier functions in the SLWF procedure, whose localization are ignored in the minimization procedure, become very delocalized. Using SLWF thus allows us to construct more localized objective Wannier functions at the expense of delocalization of the remaining ones. 

Figure \ref{wannier_1d}(c) presents the results obtained by fixing the centers of the two selectively localized orbitals to be at $x=0.5a$.  Fixing the centers increases the spread relative to the case where the centers were not constrained; however it is still much smaller than the summation of the two most localized MLWF spreads in the $J=J'=4$ case. Furthermore, not only are the centers now located at the chosen sites, but the orbitals transform as the two possible irreducible representations (even and odd parity) of the order $2$ group, respectively even though we have not forced the symmetry in any way. 

In this one-dimensional chain, we can also introduce extra Lagrange multipliers to make the orbitals transform like particular irreducible representations, as outlined in the Appendix.  For example, we force both objective Wannier functions transform as the identity about $0.5a$ in the case $J=4, J'=2$ (see Figure \ref{wannier_1d}(d)). 
The total spread has further increased relative to the previous case, with the spread of the second orbital nearly doubling, though now both orbitals are centered at $x=0.5a$ and both transform as the identity representation of the order 2 group.  Nonetheless, the largest spread of the OWF is still substantially less than the most compact MLWF.
Further insightful examples in the one dimensional chain are considered in the Appendix.

\section{Applications}
\label{section-applications}
\label{section-Computational-details}
Having demonstrated the viability of our method in simple scenarios, we now turn to realistic applications involving relevant materials. Here we study GaAs, SrMnO$_3$, and Co, as they embody three different prototypical systems. In GaAs, we will show that our method produces atomic-like orbitals of appropriate local symmetry. SrMnO$_3$ is a prototypical transition metal oxide with correlated electron properties, while in elemental Co the transition metal d-orbitals are not well separated from the less correlated s and p orbitals. 

We use the Vienna ab initio Software Package (VASP) \cite{kresse1993ab, kresse1994ab, kresse1996efficiency, kresse1996efficient} to perform DFT calculations with projector augmented wave (PAW) potentials \cite{blochl1994projector, kresse1999ultrasoft}. The exchange-correlation functional is treated within the generalized gradient approximation (GGA), as parameterized by Perdew, Burke, and Ernzerfhof (PBE) \cite{perdew1996generalized}. In all calculations, we use experimental lattice constants, which are $5.653$, $3.805$ and $3.54\AA$ for GaAs, SrMnO\textsubscript{3} and Co, respectively. The mesh of $k$-points is taken as $8\times8\times8$ with the $\Gamma$ point included. Spin polarization is not included in the calculations. All the isosurface figures are plotted using the XCrySDen program \cite{kokalj2003computer}.

While false local minima can in principle occur in our minimization procedure, they do not seem to occur in the applications presented in this section, as long as we start from reasonable trial projection functions.

\subsection{GaAs}
As shown in Ref.~\cite{marzari1997maximally} for GaAs, MLWF yields four identically localized Wannier functions (under T$_d$), exhibiting the character of $sp^3$ hybrids.  Here, as a model to test our method, we construct the same number of Wannier functions but with only one objective Wannier function, and compare the results in 3 cases: (a) constructing four Wannier functions using MLWF with $J=4, J'=4$; (b) constructing four Wannier function but using SLWF with $J=4, J'=1$; (c) constructing four Wannier function using SLWF+C, fixing the center of the objective Wannier function to be at the position of As, with $J=4, J'=1$.  In each case, the minimization is initialized with 4 trial s-orbitals, centered in the middle of the bonds as the projection functions.

Table \ref{GaAs spread comparison} reports the spreads of all four Wannier functions in each method.  As anticipated, SLWF makes the objective Wannier function (1*) most localized at the expense of the remaining Wannier functions.  SLWF thus localizes the objective Wannier function at the cost of an increased total spread summed over all four Wannier functions.  
\begin{table}
\caption{Minimized spreads in GaAs (units are $\AA^2$) from different methods. An asterisk $*$ indicates the objective Wannier function constructed in the SLWF and SLWF+C method.}
\centering
\begin{tabular}{l c c c}
\hline\hline
\hspace{2mm} & MLWF & SLWF & SLWF+C  \\[0.5ex]
\hline
\hspace{2mm}$1^*$ & 2.1977 & 1.4283 & 1.4764\\
\hspace{2mm}2 & 2.1977 & 3.0330 & 4.1243\\
\hspace{2mm}3 & 2.1977 & 3.0330 & 4.1243\\
\hspace{2mm}4 & 2.1977 & 3.0330 & 4.1243\\ [1ex]
\hline
\end{tabular}
\label{GaAs spread comparison}
\end{table}
\setlength{\tabcolsep}{2pt}
\begin{figure}
\begin{center} 
\subfigure[MLWF]{
\label{GaAs_Wannier90}   
\includegraphics[width=0.3\linewidth]{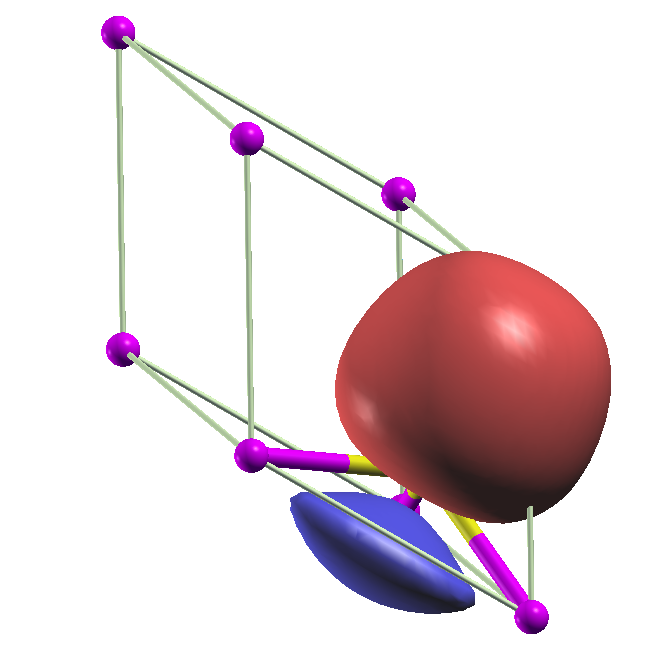}}
\subfigure[SLWF]{
\label{GaAs_selective}   
\includegraphics[width=0.3\linewidth]{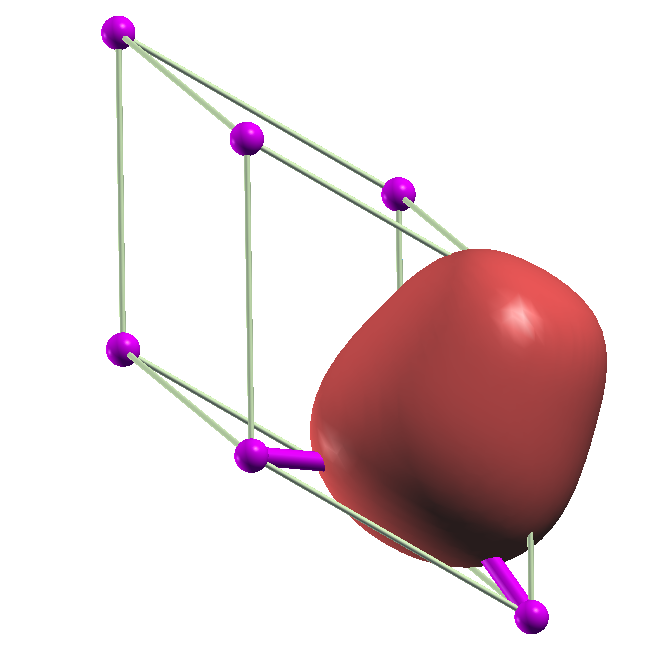}}
\subfigure[SLWF+C]{
\label{GaAs_center}
\includegraphics[width=0.3\linewidth]{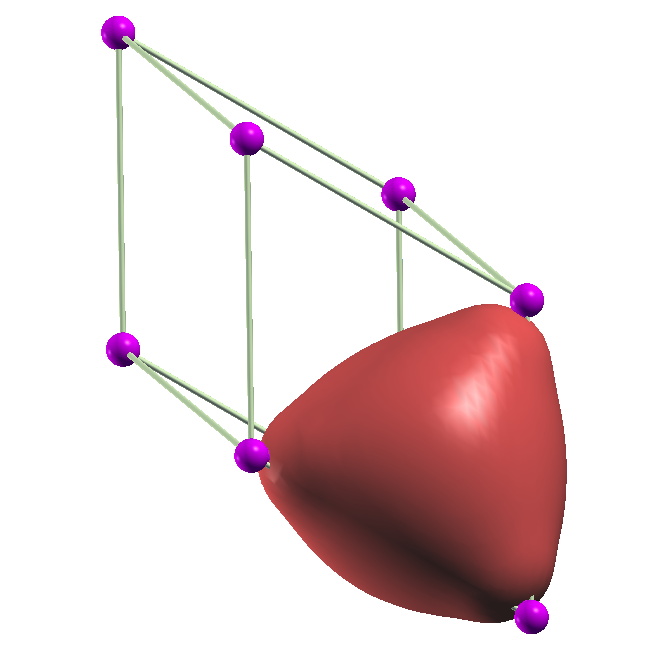}}  
\subfigure[Wannier function along the Ga-As bond.]{
\label{GaAs_bond}   
\includegraphics[width=0.9\linewidth]{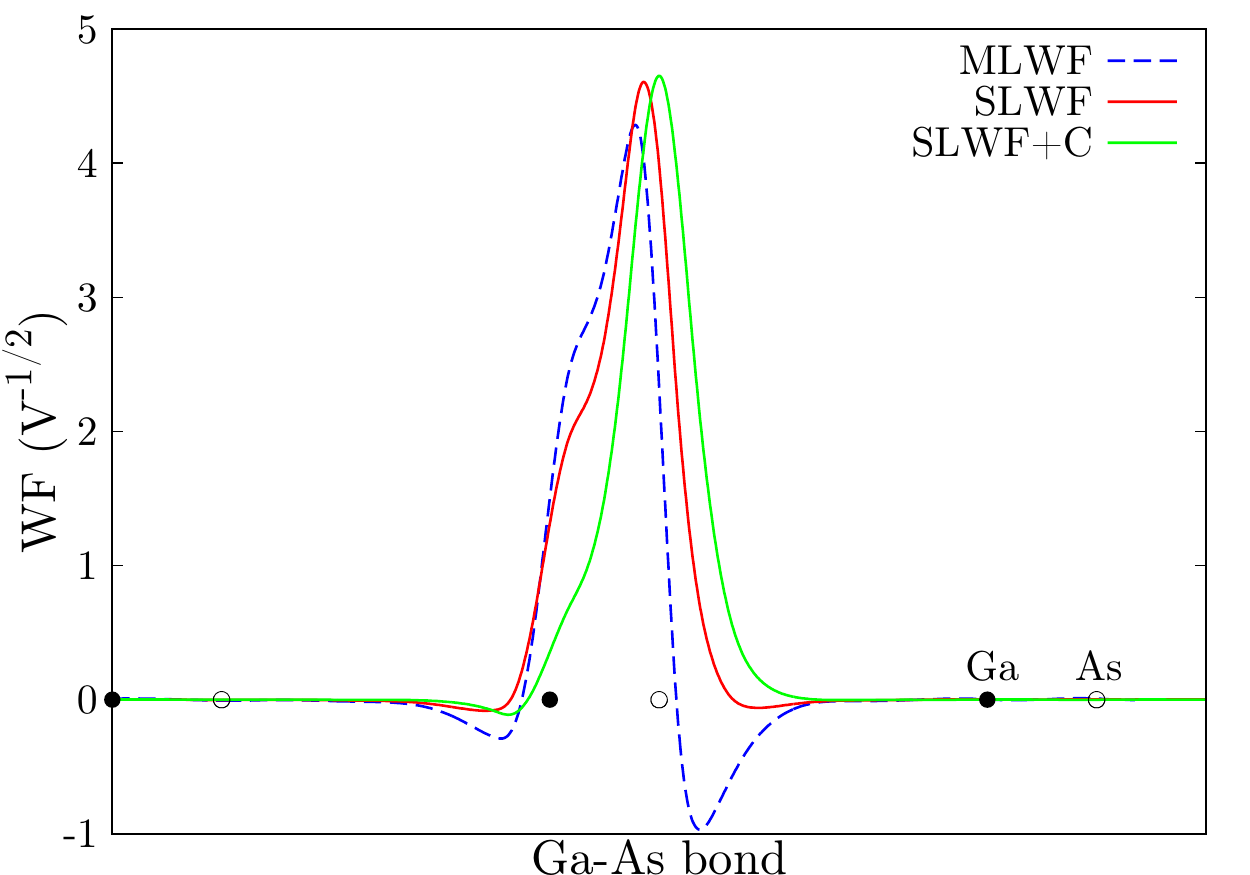}}
\caption{Wannier functions for GaAs obtained from different methods.  In panels (a), (b) and (c), Ga ions are indicated by purple spheres on the lattice corners; As ions are blocked from view by the Wannier function isosurfaces.  The absolute value of the isosurfaces is $0.5/\sqrt{V}$, where V is the unit cell volume. Isosurfaces with positive amplitudes are colored red; those with negative amplitudes are colored blue.  The Wannier functions are all real-valued. Panel (d) shows a slice of each Wannier function along the Ga-As bond. The OWF
function is plotted for the SLWF and SLWF+C methods.
}  
\label{GaAs_compare}
\end{center}
\end{figure}
SLWF also pushes the Wannier center closer to the As ion. Following Ref.~\citep{marzari1997maximally} we consider $\beta$, the ratio of the distance between the Wannier center and the Ga ion, and the length of the Ga-As bond.  Using MLWF, we obtain $\beta = 0.618$, whereas using SLWF, we obtain $\beta = 0.706$.  Of course, when we force the center to be located at the As site (SLWF+C), we have $\beta = 1$. 
In this case, fixing the center only causes a mild increase in the spread of the objective Wannier function.

In Figure \ref{GaAs_compare}, we present plots showing the objective Wannier function obtained via the different methods.  Interestingly, when we fix the center to locate at the position of As, the shape of the objective Wannier function naturally changes such that it transforms like the identity under $T_d$.  Figure \ref{GaAs_bond} further illustrates that SLWF smooths out the large bumps that arise in solutions obtained using MLWF.   

In summary, we have demonstrated that SLWF has functionality that cannot be achieved using MLWF. However, these functionalities are not clearly relevant to the physics of GaAs; the results are thus a proof of principle.

\subsection{SrMnO\textsubscript{3}}
We next turn to consider the Wannier functions corresponding to the $d$-orbitals of Mn and $p$-orbitals of O in SrMnO\textsubscript{3}. In this material, there is an isolated manifold of 14 bands, which encompasses the Fermi energy. This manifold is predominantly composed of Mn $d$ and  oxygen $p$ character (see Figure \ref{SrMnO3_bands}). MLWF will localize all 14 Wannier functions weighted equally. However, the physics of this compound is driven by correlations on the d-orbitals, so we seek a method which adequately localizes only these orbitals. We therefore apply our SLWF to localize 5 objective Wannier functions out of the total 14. We initialize using $5$  trial $d$-orbitals centered on Mn and $9$  trial $p$-orbitals centered on O as the projection functions.

\setlength{\tabcolsep}{6pt}
\begin{table}
\caption{Minimized spreads in SrMnO\textsubscript{3} and Co (units are $\AA^2$) obtained via MLWF and SLWF.}
\centering
\begin{tabular}{|c | c c || c c |}
\hline
&\multicolumn{2}{c||}{SrMnO$_3$} & \multicolumn{2}{c|}{Co} \\
  & MLWF & SLWF & MLWF & SLWF \\[0.5ex]
\hline
$3z^2-r^2$ & 0.5056 & 0.5006 &0.5144 & 0.5051  \\
$xz$ & 0.5486 & 0.5467 & 0.8505 & 0.5615 \\
$yz$ & 0.5486 & 0.5467 & 0.8505 & 0.5615 \\
$x^2-y^2$ & 0.5056 & 0.5006  & 0.5144 & 0.5051 \\
$xy$ & 0.5486 & 0.5467 & 0.8505 & 0.5615 \\ [1ex]
\hline
\end{tabular}
\label{SrMnO3 spread comparison}
\end{table}

In Table \ref{SrMnO3 spread comparison}, we compare the spreads of the 5 $d$-like MLWFs and the OWFs.  Though the differences are very minimal, SLWF constructs $d$-like Wannier functions that are slightly more localized. 
\begin{figure*}
\begin{center} 
\subfigure[MLWF.]{
\label{SrMnO3_compare2}   
\includegraphics[width=0.19\linewidth]{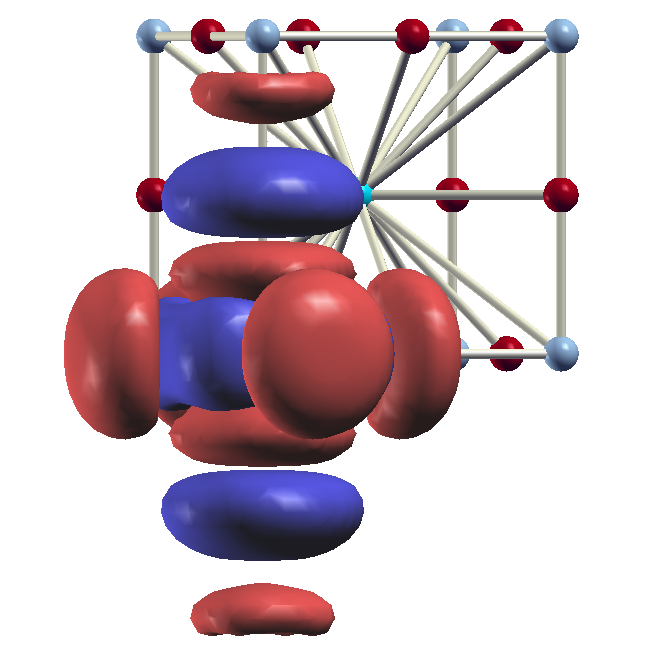}\includegraphics[width=0.19\linewidth]{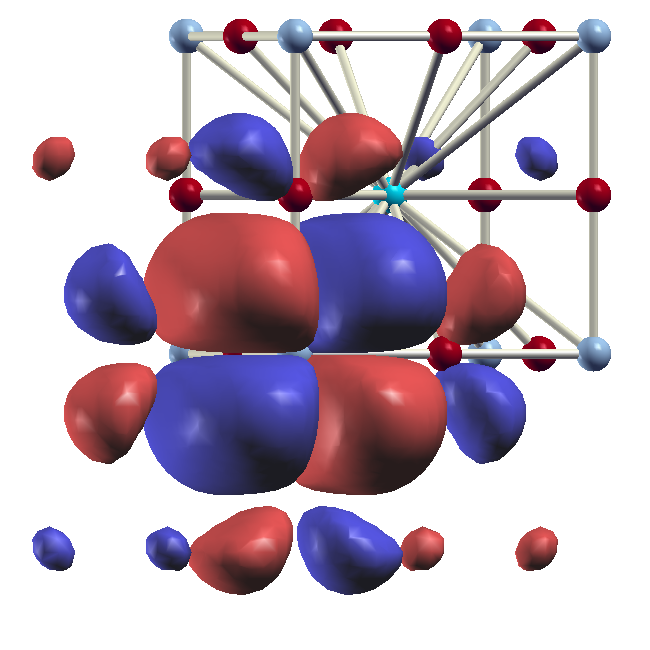}\includegraphics[width=0.19\linewidth]{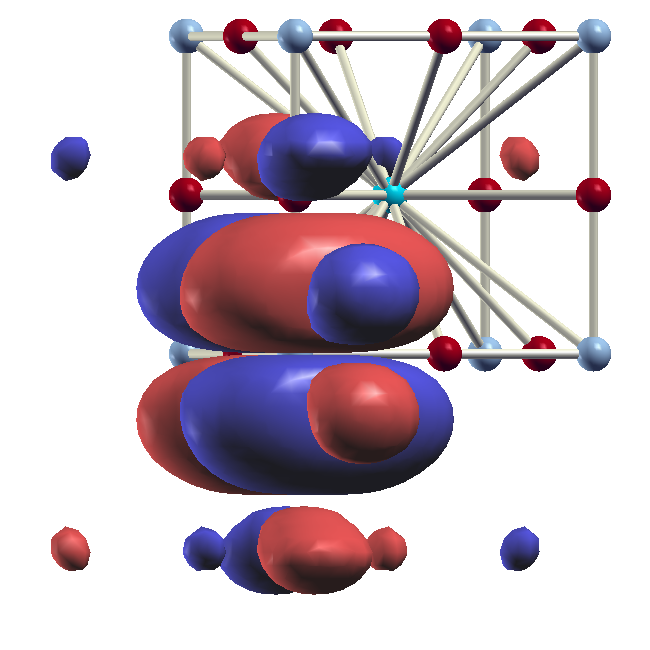}\includegraphics[width=0.19\linewidth]{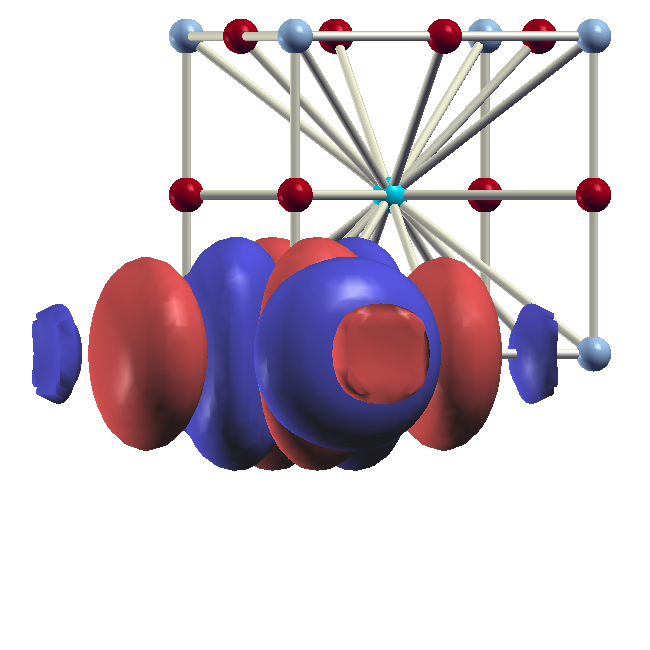}\includegraphics[width=0.19\linewidth]{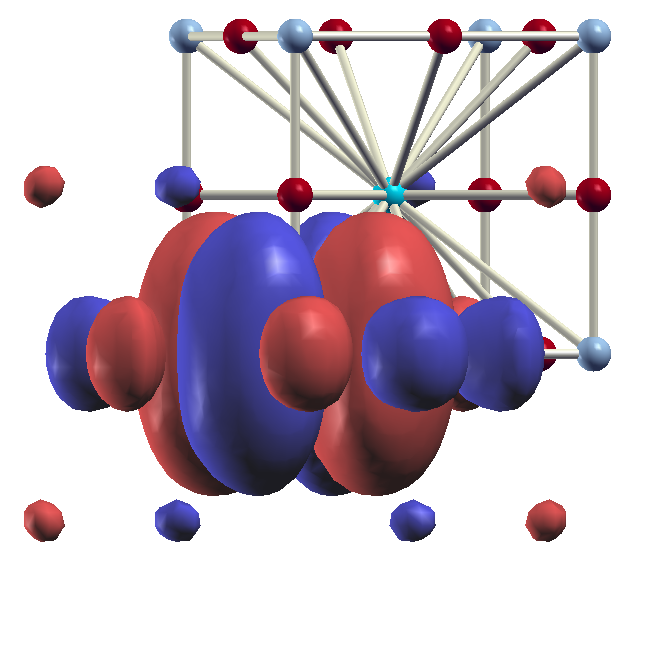}}
\subfigure[SLWF.]{
\label{SrMnO3_compare1}   
\includegraphics[width=0.19\linewidth]{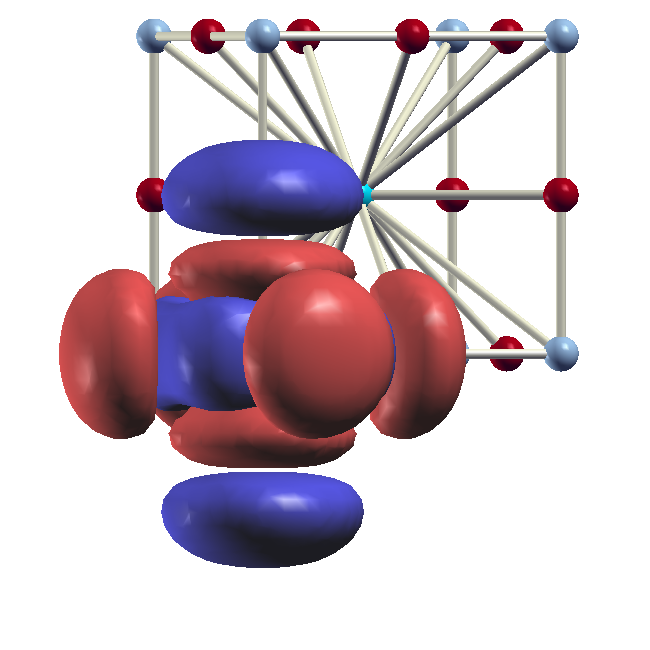}\includegraphics[width=0.19\linewidth]{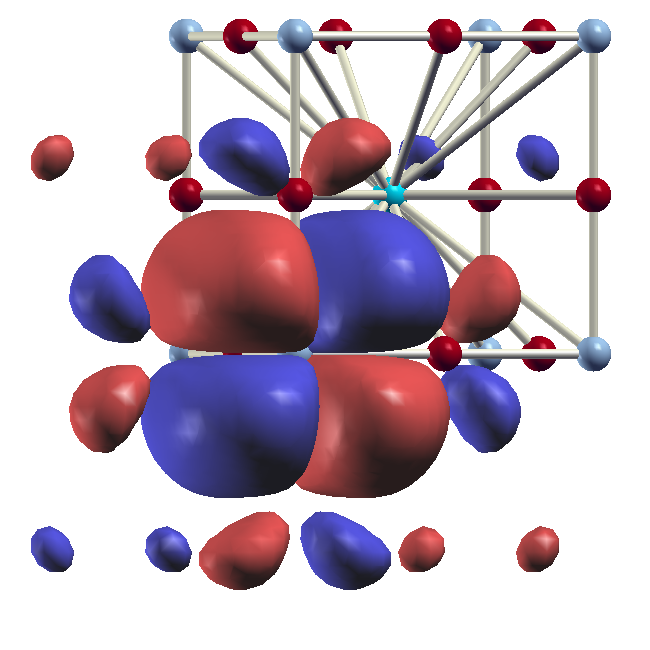}\includegraphics[width=0.19\linewidth]{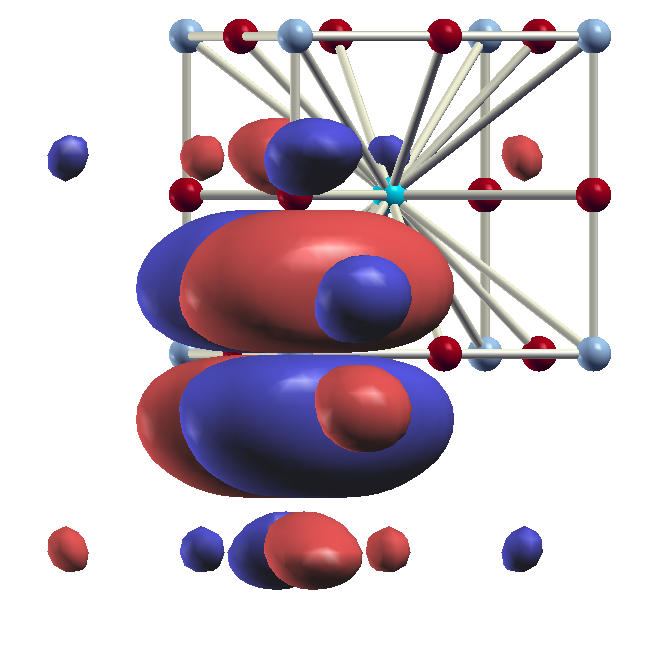}\includegraphics[width=0.19\linewidth]{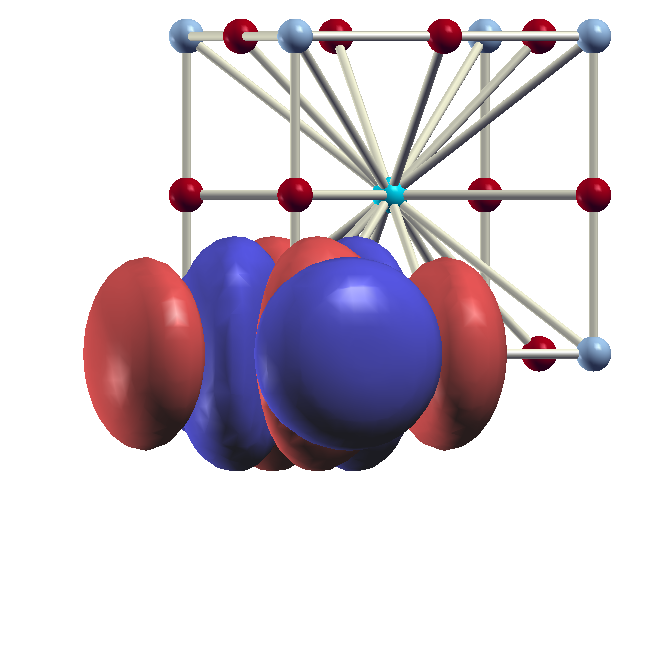}\includegraphics[width=0.19\linewidth]{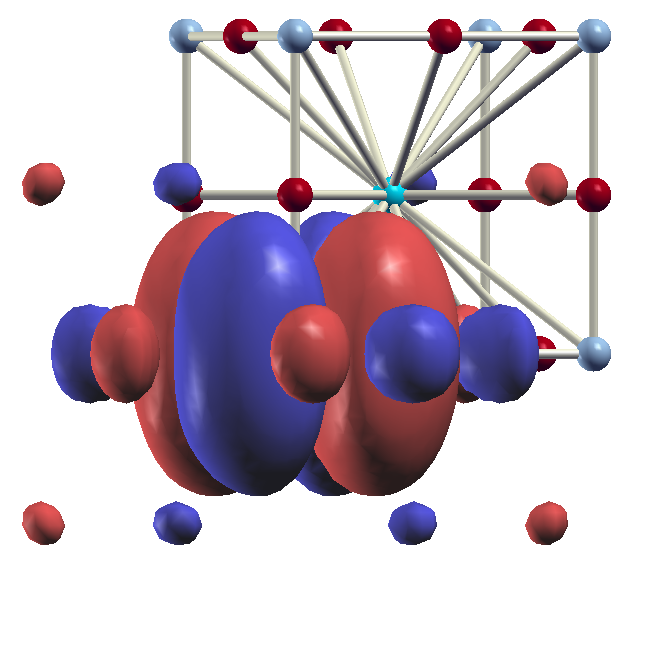}} 
\caption{Comparison between the MLWF and SLWF method for SrMnO\textsubscript{3}. Sr is the light blue sphere in the center, Mn is the blue sphere on the corner, O is the red sphere. In both panels, the 5 d-like Wannier functions are in the order of $3z^2-r^2$, $xz$, $yz$, $x^2-y^2$, $xy$, from left to right. The absolute value of the isosurfaces  is $0.1/\sqrt{V}$.}
\label{SrMnO3_compare}
\end{center}
\end{figure*}
\begin{figure}
\begin{center}
\includegraphics[width=0.9\columnwidth]{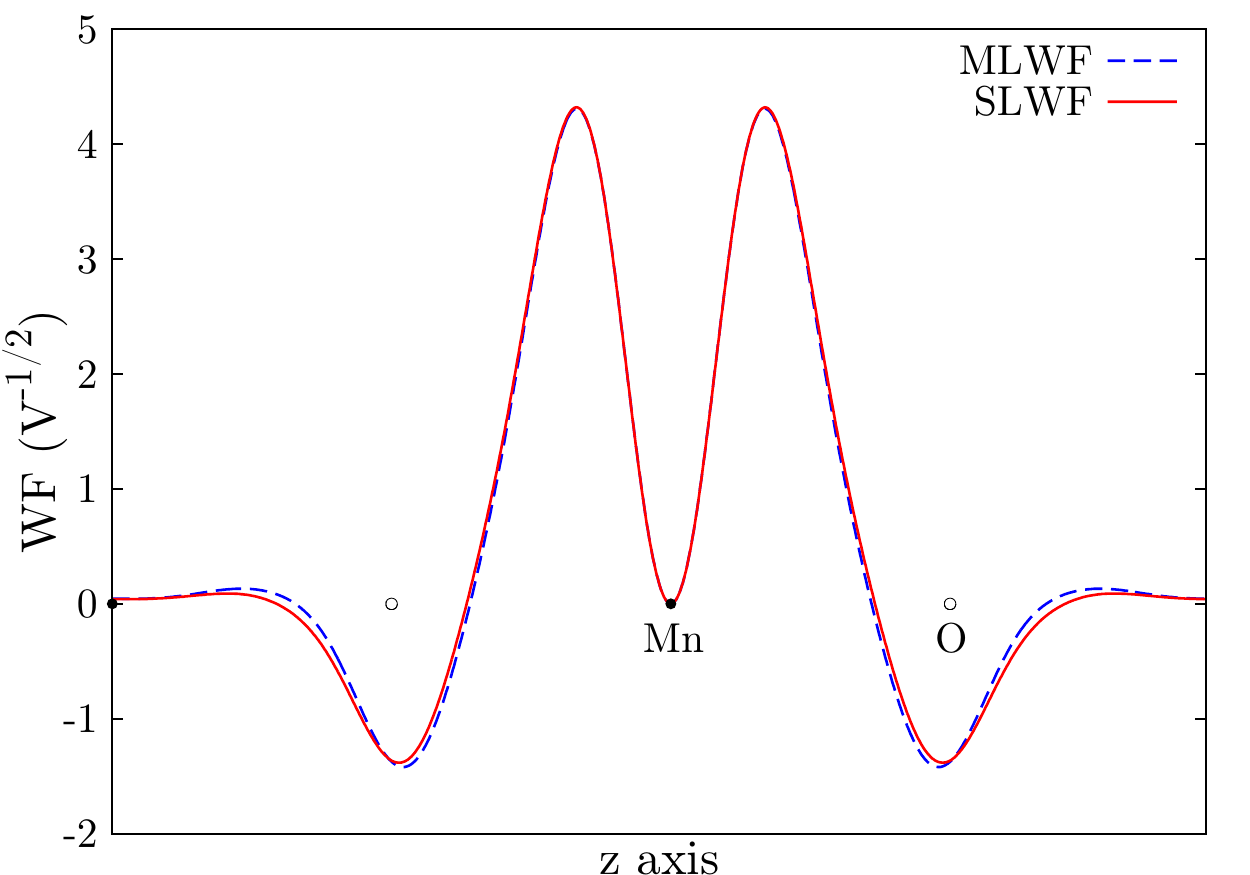}
\caption{Wannier function along the $z$ axis for the $3z^2-r^2$-like Wannier function generated from MLWF
and SLWF in SrMnO\textsubscript{3}.}
\label{SrMnO3_compare_axis}
\end{center}
\end{figure}
In Figure \ref{SrMnO3_compare}, we plot an isosurface of $0.1/\sqrt{V}$ for both cases, in which the Wannier functions transform like $e_g$ and $t_{2g}$ orbitals centered at the Mn site. Compared with the MLWFs, the OWFs show noticeably smaller tails. However, since the value of the isosurface in the plot is quite small, the differences are in fact very minimal. This feature is also apparent in the plot along the $z$ axis for the $3z^2-r^2$-like MLWF and OWF shown in Figure \ref{SrMnO3_compare_axis}.

In summary, in the high symmetry, separated band case of SrMnO$_3$, our SLWF procedure has very minimal differences as compared to MLWF in terms of the spread for this test case. Further analysis will be performed in the next section where we analyze the Hamiltonian.

\subsection{Co}
\label{section-Co}
\setlength{\tabcolsep}{2pt}

\begin{figure*}
\begin{center} 
\subfigure[MLWF.]{
\label{Co_Wannier90}   
\includegraphics[width=0.19\linewidth]{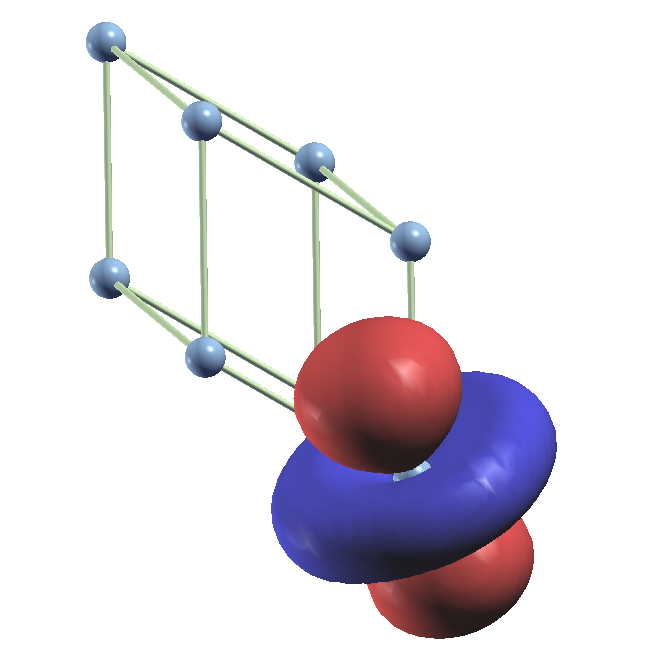}
\includegraphics[width=0.19\linewidth]{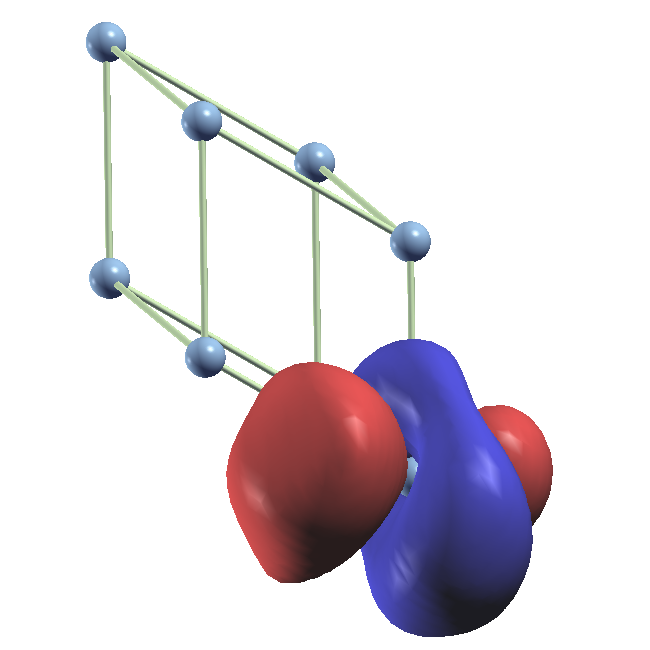}
\includegraphics[width=0.19\linewidth]{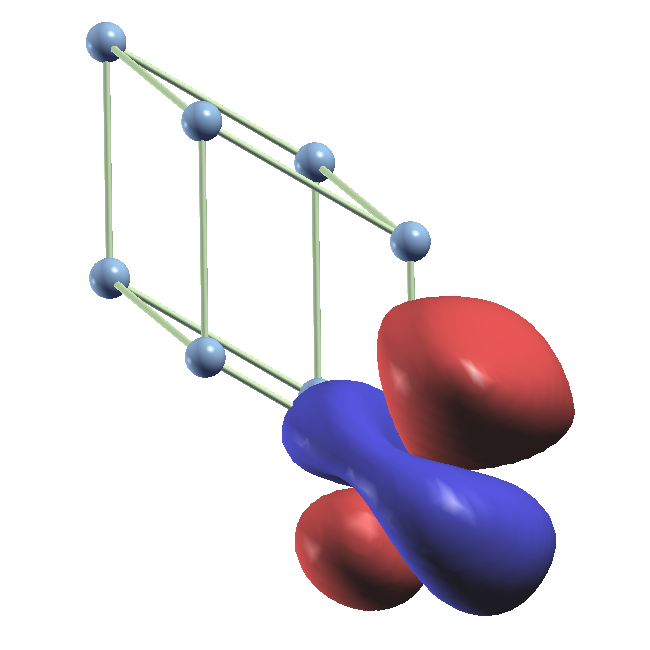}
\includegraphics[width=0.19\linewidth]{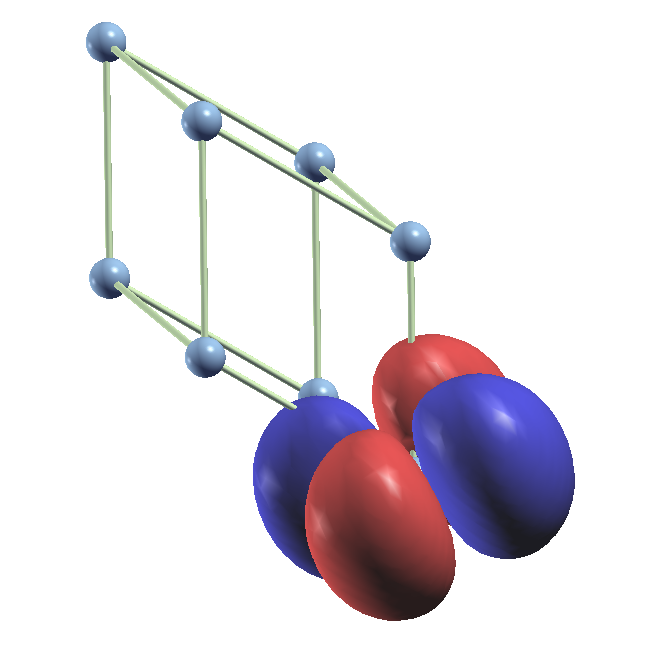}
\includegraphics[width=0.19\linewidth]{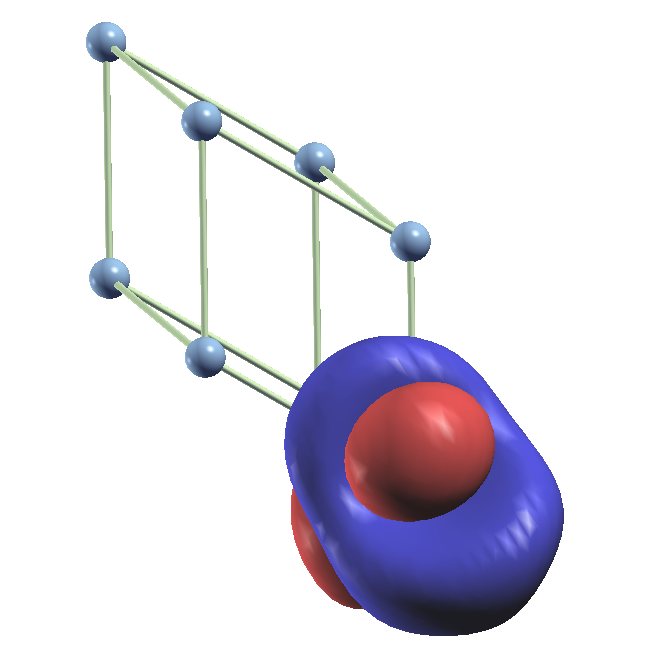}}
\subfigure[SLWF.]{
\label{Co_selective}   
\includegraphics[width=0.19\linewidth]{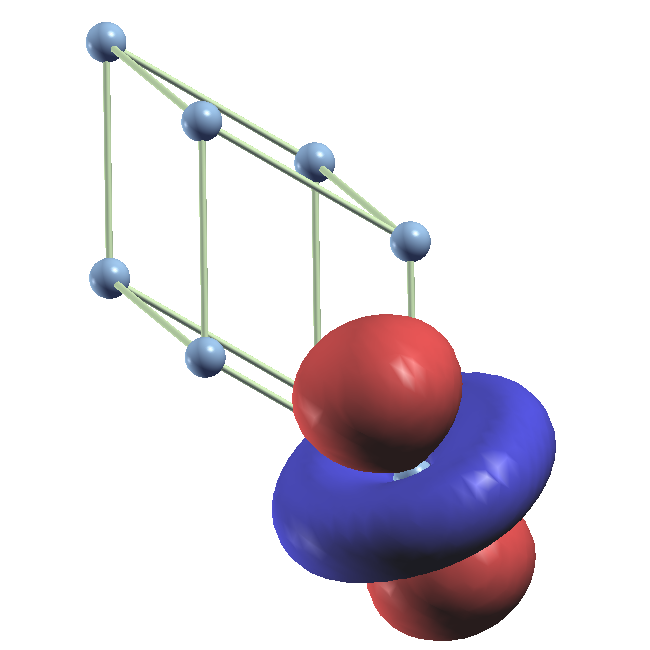}\includegraphics[width=0.19\linewidth]{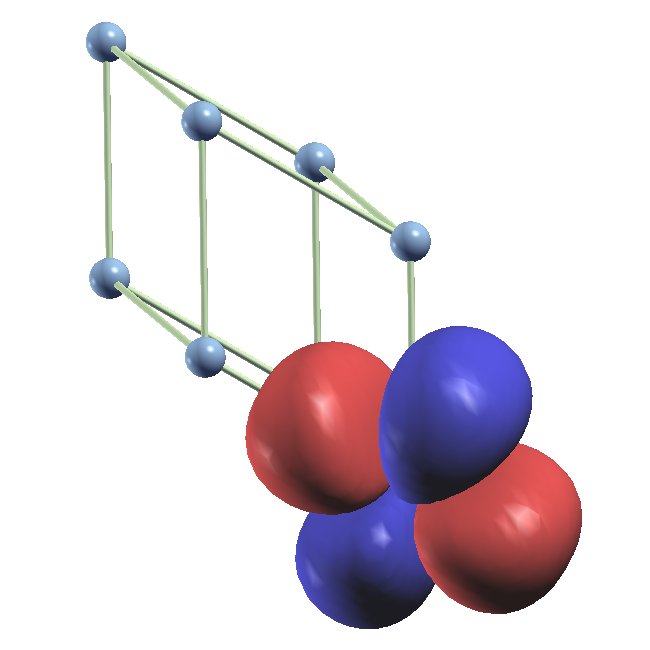}\includegraphics[width=0.19\linewidth]{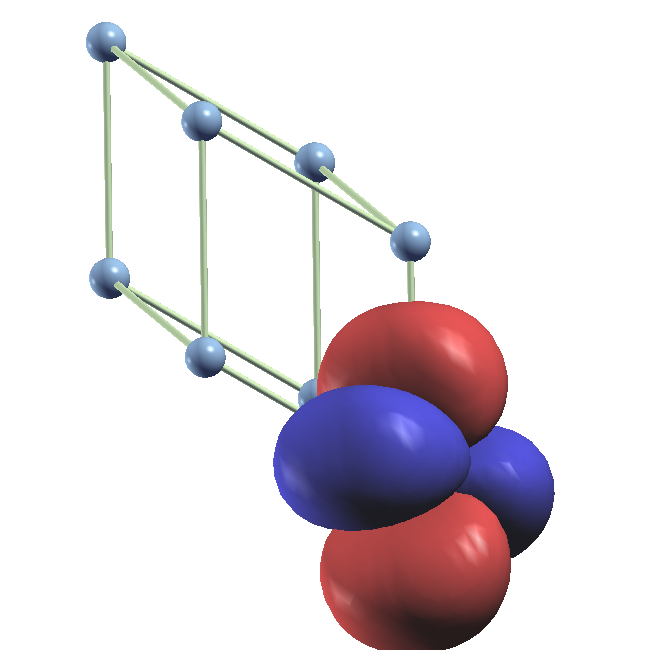}\includegraphics[width=0.19\linewidth]{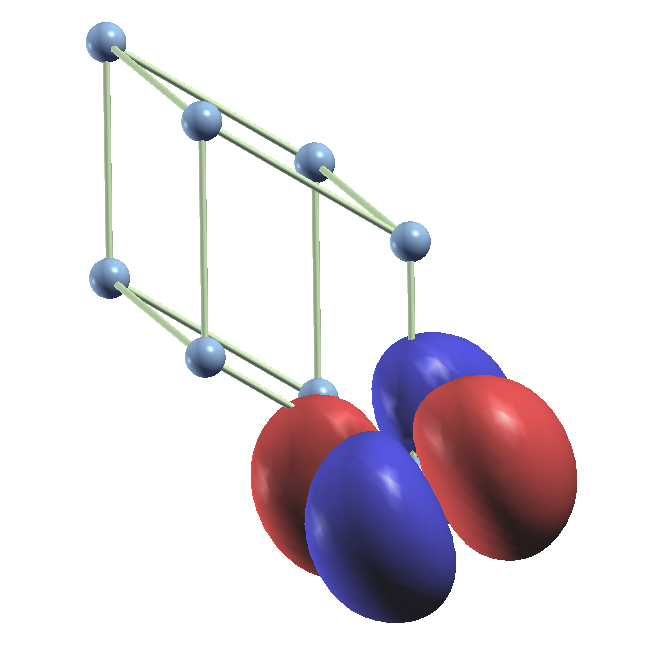}\includegraphics[width=0.19\linewidth]{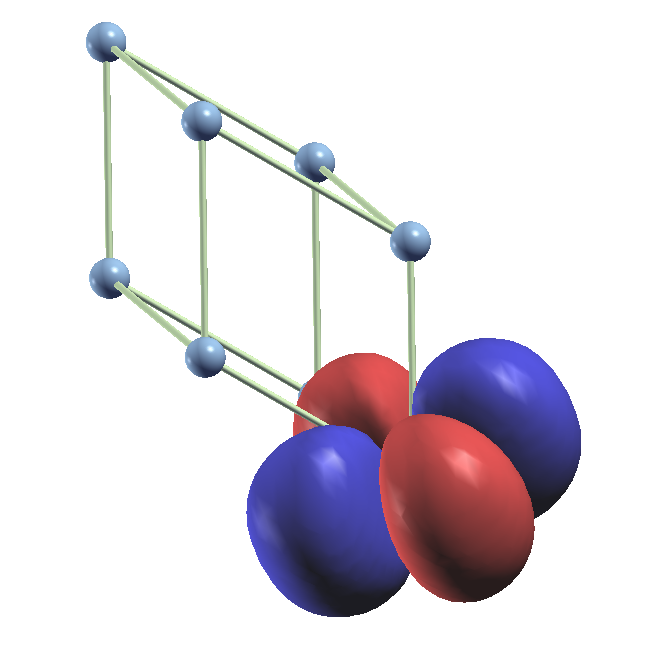}}  
\caption{Comparison between the MLWF and SLWF method for Co. Co atoms are indicated by blue spheres on the lattice corners. In both panels, the 5 d-like Wannier functions are in the order of $3z^2-r^2$, $xz$, $yz$, $x^2-y^2$, $xy$, from left to right. The absolute value of the isosurfaces is $0.2/\sqrt{V}$.}
\label{Co_compare}
\end{center}
\end{figure*}

\begin{figure}
\begin{center}
\includegraphics[width=0.9\columnwidth]{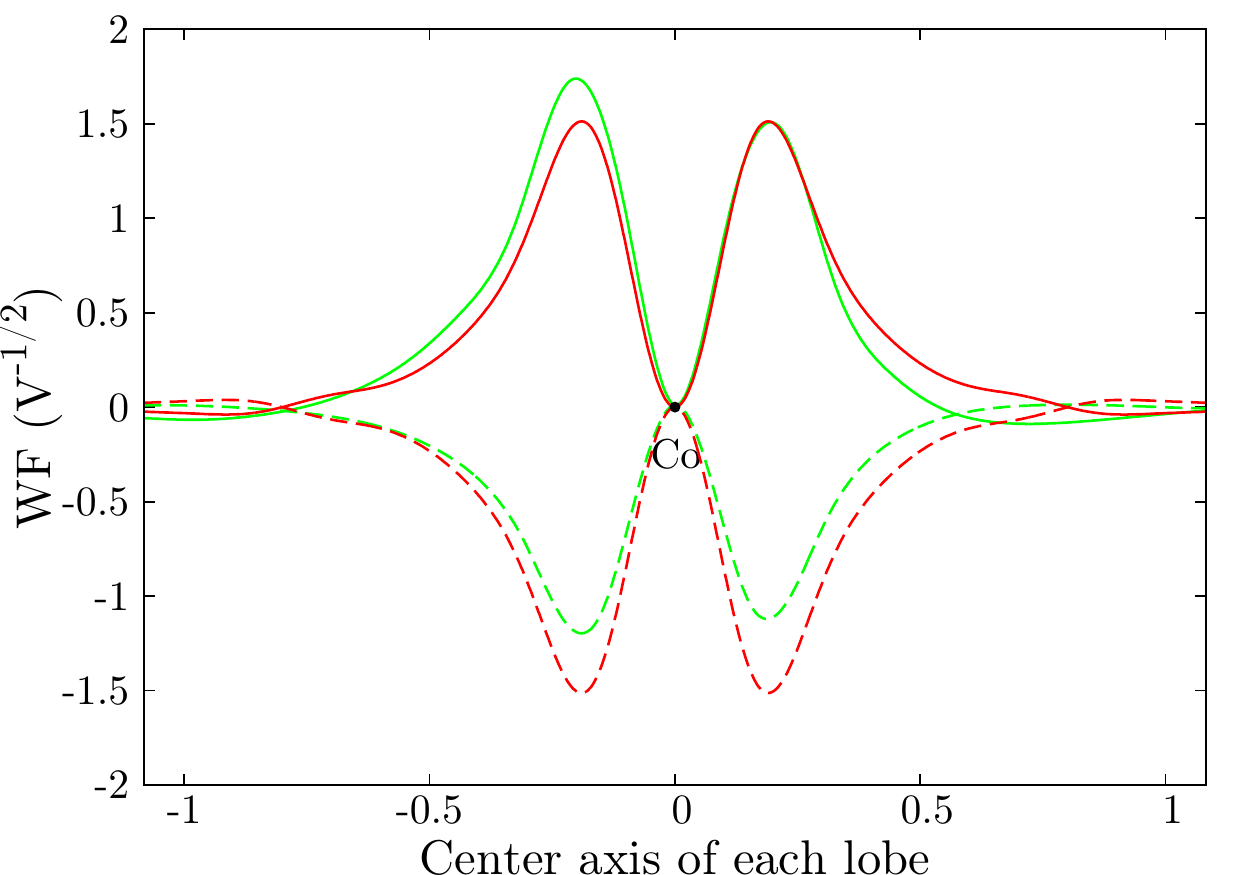}
\caption{Wannier function along the center axis of each lobe for the $xz$-like Wannier function generated from MLWF and SLWF in Co. The light lines (green online) represent MLWF while the dark lines (red online) represent SLWF. For a given method, the solid lines are along the lobes with positive phase while the dashed lines are along the ones with negative phase. The horizontal axis is in units of the lattice constant. }
\label{Co_compare_axis}
\end{center}
\end{figure}

The final example considered is Co, a transition metal in which  the electronic structure is less ionic than a transition metal oxide. Additionally, and unlike the cases considered in the previous two subsections,  there is no clear separation between the highest retained bands and subsequent bands in the electronic structure. Specifically, we consider Co in the face centered cubic (FCC) structure, and we construct a set of 6 Wannier orbitals by including the lowest 6 bands, which encompasses the narrow $d$ bands and free-electron-like $s$ band which hybridizes with the $d$ bands. In our SLWF construction we choose $5$ objective Wannier functions (ie. $J=6$ and $J^\prime=5$). We initialize both cases by using atom-centered trial $d$-orbitals together with a  trial $s$-orbital which is centered around one of the tetrahedral-interstitial sites.

The right-hand column of Table \ref{SrMnO3 spread comparison} summarizes the spreads of the 5 d-like Wannier functions obtained using the two localization methods. As shown, SLWF decreases the spread relative to MLWF in all $5$ orbitals, with strong
decreases in the $t_{2g}$ manifold. It is also worth noting that the OWFs we obtain using SLWF in Co are nearly as localized as in SrMnO$_3$.  In Figure \ref{Co_compare}, we compare the isosurface plots of the $d$-like MLWFs and the OWFs. In the MLWF case we still use the same orbital labels for the three $t_{2g}$-like orbitals for simplicity although the $t_{2g}$-like MLWFs have apparently lost $t_{2g}$ symmetry and there are indications that the $s$-orbital has been mixed in. In Figure \ref{Co_compare_axis}, we plot one of the $t_{2g}$-like MLWFs and OWFs along the center axis of each lobe, in units of the lattice constant. It is obvious that the 4 lobes in MLWF do not have the same shape any more while the OWF preserves the $t_{2g}$ symmetry. Therefore, we conclude that SLWF offers improvements for creating atomic-like $d$-orbitals for use in beyond-DFT methods in transition metals.

\section{Hamiltonian in MLWF and SLWF bases}
Thus far we have only examined the spatial properties of the Wannier orbitals, but another important aspect of the Wannier orbitals
is the nature of the Hamiltonian in this basis.  In order to elucidate this and to understand the differences in the Hamiltonians for MLWF and SLWF, we will follow the analysis put forward by Toropova et.~al.~\cite{toropova_bands} in constructing Hamiltonians for SrMnO$_3$ and Co.  In general we will have a $k$-space Hamiltonian with an objective orbital block and some other block of states that hybridizes with the objective block. In our  test cases the objective orbitals correspond to a $d$-block while the hybridizing orbital would be an $s$-orbital for Co and $p$-orbitals for SrMnO$_3$:
\begin{eqnarray}
\label{Ham}
H(\bf{k})=\left( \begin{array}{cc}
H_d(\bf{k}) & V(\bf{k})\\
V^{\dagger}(\bf{k}) & H_{sp}(\bf{k}) \end{array} \right)
\end{eqnarray}
where the subscript $sp$ simply denotes the block of orbitals that are not $d$.
\begin{figure}
\begin{center}    
\subfigure[MLWF.]{
\label{SrMnO3_bands_MLWF}
\includegraphics[width=0.9\columnwidth]{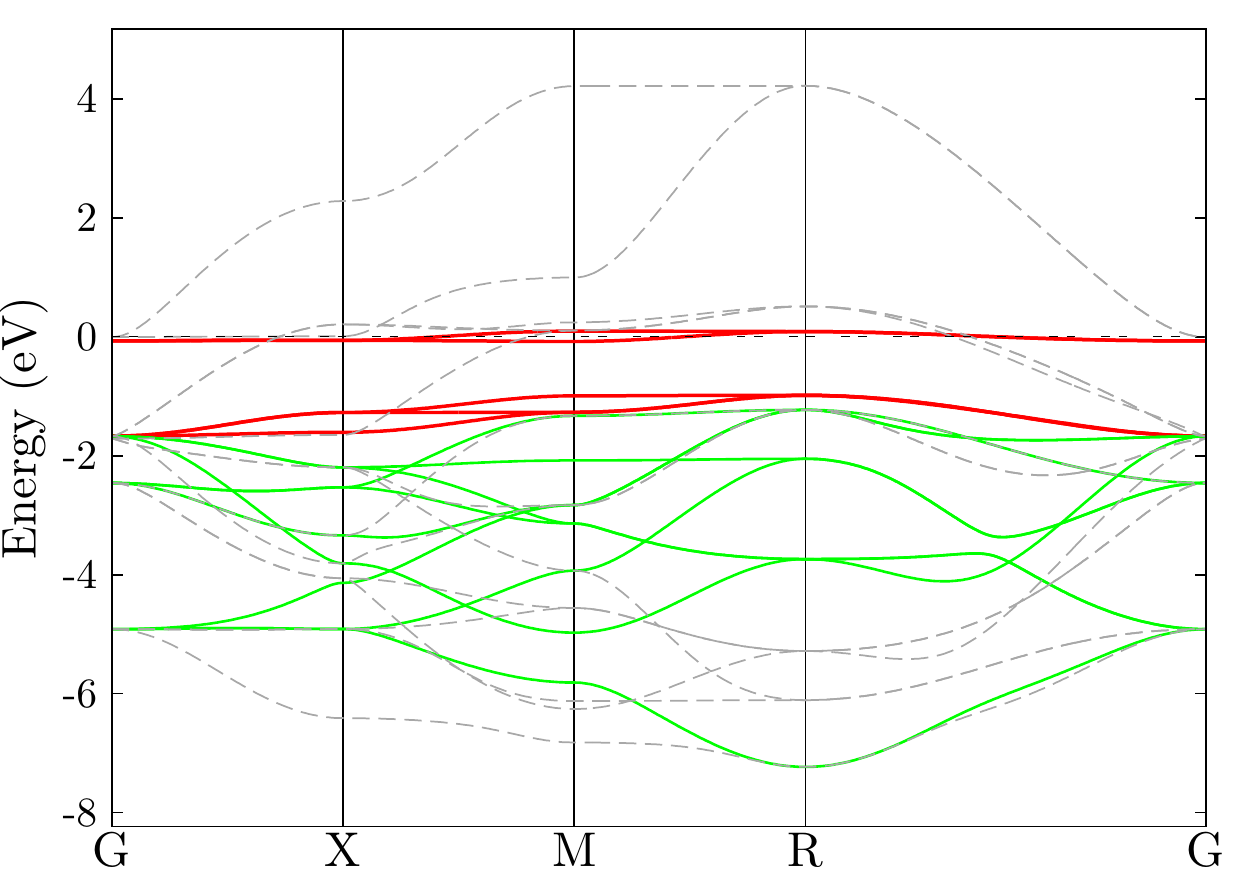}}
\subfigure[SLWF.]{
\label{SrMnO3_bands_SLWF}
\includegraphics[width=0.9\columnwidth]{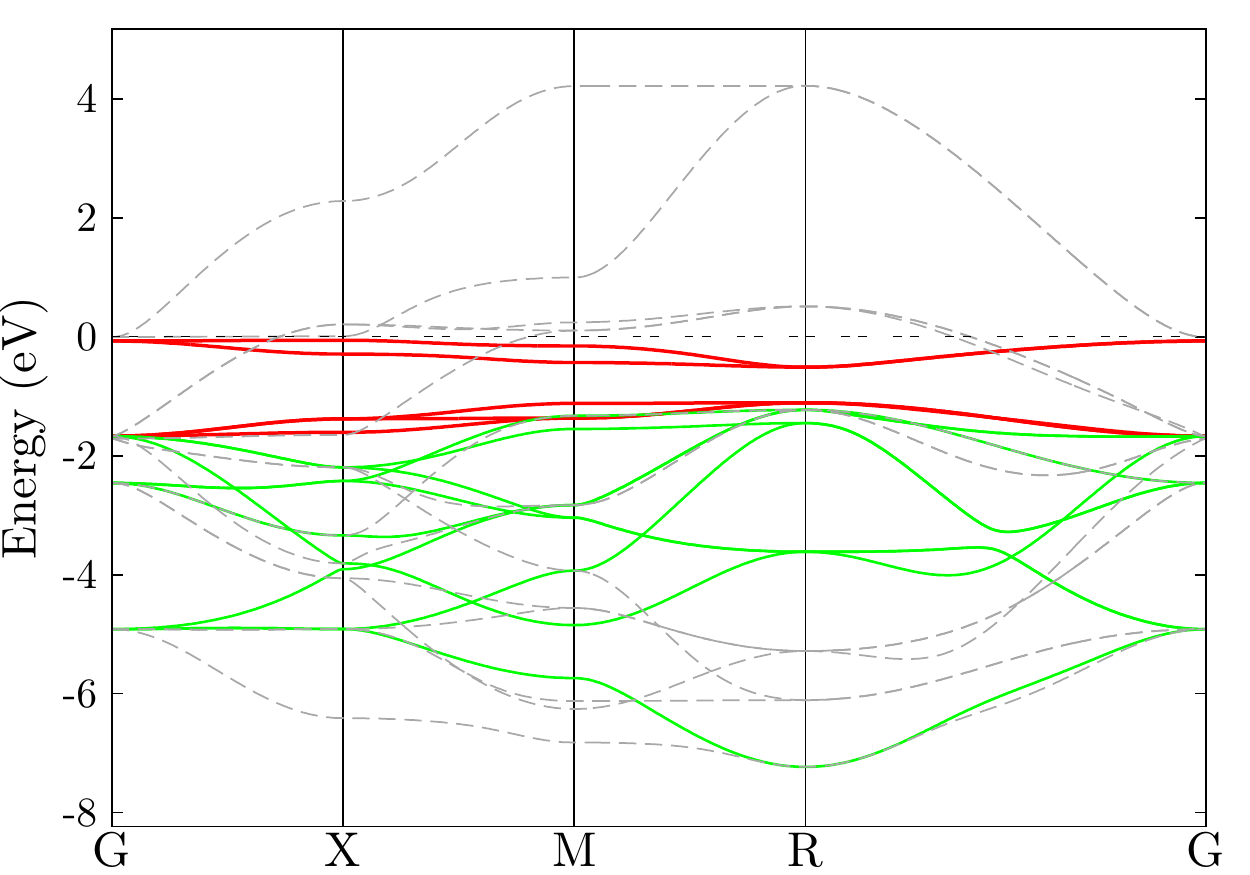}}
\caption{Sliced Band structures of SrMnO$_3$. In both panels, dashed gray lines represent the DFT band structure; solid dark lines (red online) represent bands for the block $H_d$; solid light lines (green online) represent bands for the block $H_p$. }
\label{SrMnO3_bands}
\end{center}
\end{figure}
\begin{figure}
\begin{center}    
\subfigure[MLWF.]{
\label{Co_bands_MLWF}
\includegraphics[width=0.9\columnwidth]{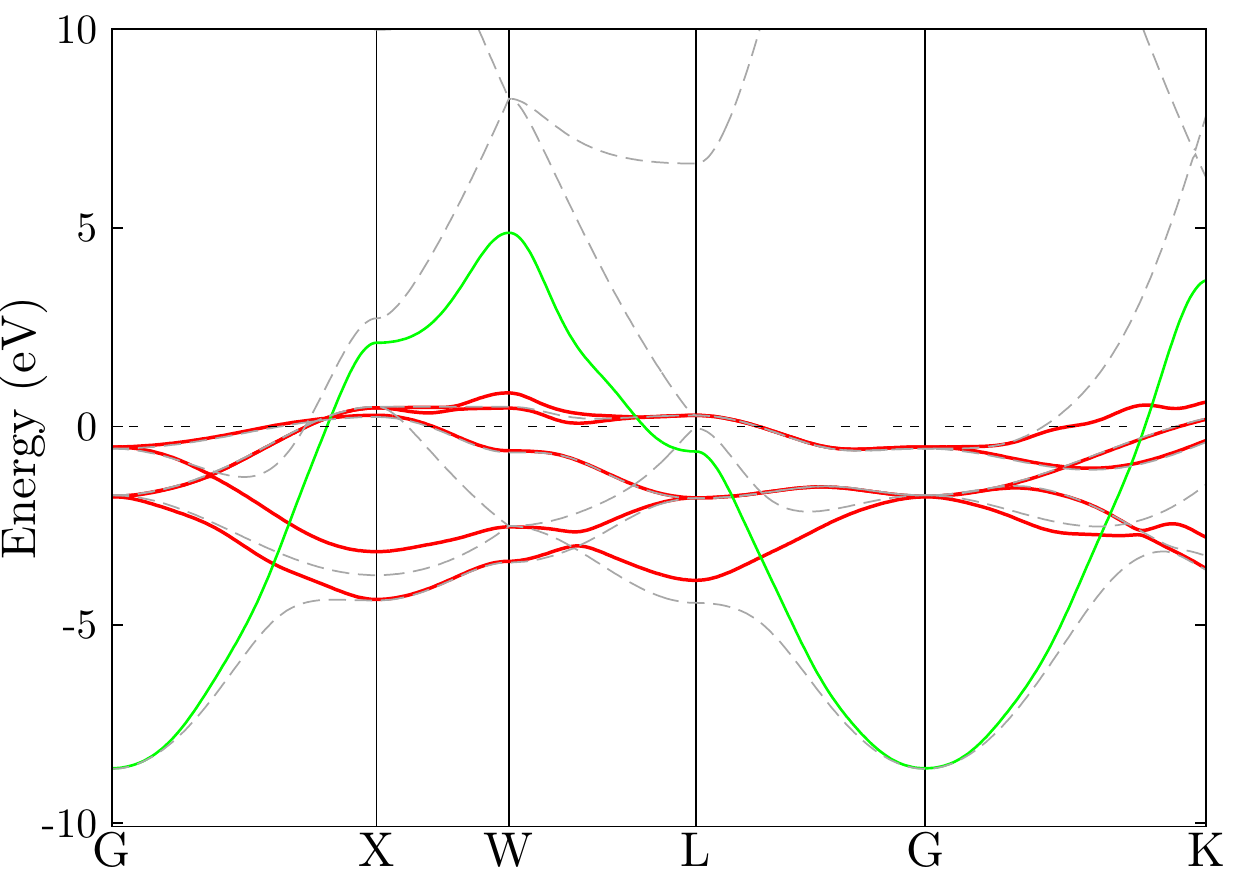}}
\subfigure[SLWF.]{
\label{Co_bands_SLWF}
\includegraphics[width=0.9\columnwidth]{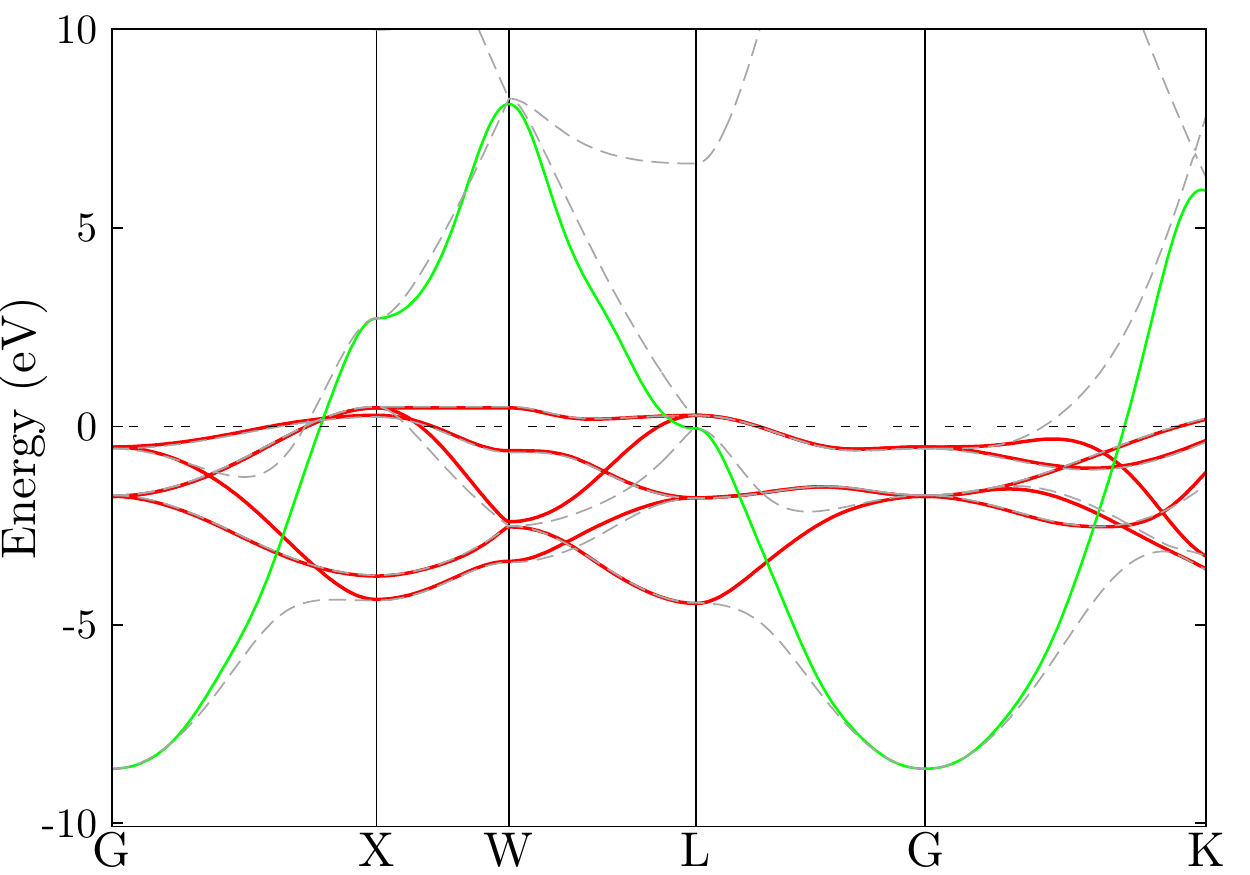}}
\caption{Sliced Band structures of Co. In both panels, dashed gray lines represent the DFT band structure; solid dark lines (red online) represent bands for the block $H_d$; solid light lines (green online) represent bands for the block $H_s$.
}
\label{Co_bands}
\end{center}
\end{figure}
The effect of hopping within the $d$-manifold versus hybridization can easily be seen in the ``sliced" band structure. This simply
amounts to zeroing $V(k)$ and then diagonalizing the separate blocks of the Hamiltonian at each $k$-point, yielding a set of bands for each block. This allows one to see the bandwidth generated solely from hopping within the respective manifold, and the difference with the DFT band structure indicates the role of hybridization. We consider the case of SrMnO$_3$ and Co, following the exact same Wannier procedure as was outlined above.

In the case of SrMnO$_3$, we see that the $d$-bands have several differences (see Figure \ref{SrMnO3_bands}). The $t_{2g}$ bands are narrower for SLWF, while the  $e_g$ bands are narrower for MLWF. The latter observation is particularly counterintuitive given that one clearly observes an enhanced localization of the objective $e_g$ orbitals in Figure \ref{SrMnO3_compare}. 
However, these differences are most likely not relevant for actual calculations.

In the case of Co, the differences between SLWF and MLWF are larger (see Figure \ref{Co_bands}). In the case of MLWF, one can see that at the $W$ point the $s$-band is roughly $3eV$ away from the DFT band, indicating a strong hybridization between the 
$s$-like and $d$-like MLWFs must be present. This is a symptom of the character mixture that we visually observed in Figure \ref{Co_compare} and quantified via the substantially larger spread of the $t_{2g}$-like orbitals. In contrast, the $s$-band obtained in the SLWF procedure nicely tracks the DFT band in the region of the $W$-point. The complimentary aspect of this result is that the sliced SLWF $d$-bands more closely track the relevant DFT bands. The same points can be made near the $K$-point. It appears clear in this case that the SLWF procedure results in a more appropriate set of $d$-orbitals from the perspective of the Hamiltonian.

\section{Conclusions}
We have generalized  the algorithm introduced by Marzari and Vanderbilt \cite{marzari1997maximally} to allow for the maximal
localization of a subset of Wannier functions, with fixed centers and symmetry. This scheme allows us to achieve greater localization for the selected subset of Wannier functions. We found that simply fixing the Wannier center produced orbitals that transformed as appropriate irreducible representations of the local point group even without specifying the symmetry. 

We illustrated our method on GaAs, SrMnO$_{3}$, and Co. From the study of GaAs we demonstrate the power of our approach by constructing a single Wannier orbital which transforms like the identity, in  addition to three delocalized orbitals which span the $4$-band $s-p$ manifold.  In the case of SrMnO$_3$, we found that our SLWF procedure yielded results very similar to those found in the MLWF procedure,  suggesting that MLWF may be sufficient for beyond-DFT calculations in transition metal oxides. In the
case of Co, SLWF offer notable improvements and results in a much purer set of $d$-orbitals, which could be very important in the context of DFT+DMFT calculations. Future work should be performed explicitly comparing these two approaches in the context of DFT+DMFT. Implementing our approach within existing MLWF codes is straightforward.

{\bf Acknowledgments.} We thank Andreas Kl\"ockner and Se Young Park for helpful discussions.  AJM was supported by DOE-ER046169.
CAM, RW, EAL, and HP were funded by NSF under contract DMR-1122594.

\bibliographystyle{apsrev4-1.bst}
\bibliography{refs}

\appendix

\section*{Appendix: Fixing symmetries \label{symmetries}}
In this Appendix we show how we enforce symmetry constraints in one-dimensional systems. Assuming that the center of symmetry is at $x_{0}$, the functional we want to minimize is:
\begin{eqnarray}
\label{sym_objective}
\Omega_s &=& \Omega_c + \lambda_s \sum_{n=1}^{J_s}\int_{-\infty}^{\infty} |(1-\sigma_{x_0})w_n(x)|^2 dx \nonumber \\
&+&\lambda_s\sum_{n=J_s+1}^{J'}\int_{-\infty}^{\infty}  |(1+\sigma_{x_0})w_n(x)|^2 dx
\end{eqnarray}
where $J_s$ is the number of the objective Wannier functions we would like to be symmetric, while the other $J'-J_s$ would be antisymmetric; $\lambda_s$ is a Lagrange multiplier for the corresponding constraint; $\sigma$ is the mirror operator. 

We define:
\begin{eqnarray}
I_{mn}^k&=&\frac{1}{N}\sum_{k'}\left[\int_{-\infty}^{\infty}\psi_{nk}(x) \sigma_{x_0}\psi_{mk'}^*(x)dx\right.\nonumber \\
&&\left.+\int_{-\infty}^{\infty} \psi_{mk'}^*(x)\sigma_{x_0}\psi_{nk}(x)dx\right]
\end{eqnarray}
Under the infinitesimal unitary transformation, we have:
\begin{eqnarray}
&&d\left[\int_{-\infty}^{\infty} |(1\pm \sigma_{x_0})w_n(x)|^2 dx\right]\nonumber\\
&=&\pm\frac{2}{N}\sum_{k}\sum_{m=1}^J\operatorname{Re}(I_{nm}^k dW_{mn}^{k})
\end{eqnarray}
Thus,
\begin{eqnarray}
d\Omega_{s}&=&d\Omega_{c}-\lambda_s\frac{2}{N}\sum_{k}\left[\sum_{n=1}^{J_s}\sum_{m=1}^J\operatorname{Re}(I_{nm}^k dW_{mn}^{k})\right.\nonumber \\
&&\left.-\sum_{n=J_s+1}^{J'}\sum_{m=1}^J\operatorname{Re}(I_{nm}^k dW_{mn}^{k})\right]\nonumber\\
&\label{whatever7}
\end{eqnarray}
The gradient of $\Omega_s$ is then:
\begin{eqnarray}
G_{s,mn}^{k}&=&\frac{d\Omega_s}{dW_{nm}^k}\\
&=&G_{c,mn}^{k}+\nonumber\\
&&\begin{cases}
-\lambda_s\left( I_{mn}^k-I^{k*}_{nm}\right),&m\leq J_s,n\leq J_s\\
-\lambda_s\left(I_{mn}^{k}+I_{nm}^{k*}\right),&m\leq J_s,J_s<n\leq J'\\
-\lambda_s I_{mn}^{k},& m\leq J_s,n>J'\\
\lambda_s\left(I_{mn}^{k}+I_{nm}^{k*}\right), & J_s<m\leq J',n\leq J_s\\
\lambda_s\left(I_{mn}^{k}-I_{nm}^{k*}\right), & J_s<m\leq J',J_s<n\leq J'\\
\lambda_s I_{mn}^k, & J_s<m\leq J',n>J'\\
\lambda_s I^{k*}_{nm}, & m>J',n\leq J_s\\
-\lambda_s I^{k*}_{nm},& m>J',J_s<n\leq J'\\
0, &  m>J',n>J'\nonumber
\end{cases}
\end{eqnarray}
Using this method, we can ensure that the objective Wannier functions preserve arbitrary symmetries in a one-dimensional system in addition to maintaining fixed centers, all while maintaining a high degree of localization by performing  selective localization.    

\begin{figure}[htb]
\begin{center}
\includegraphics[width=\linewidth,clip= ]{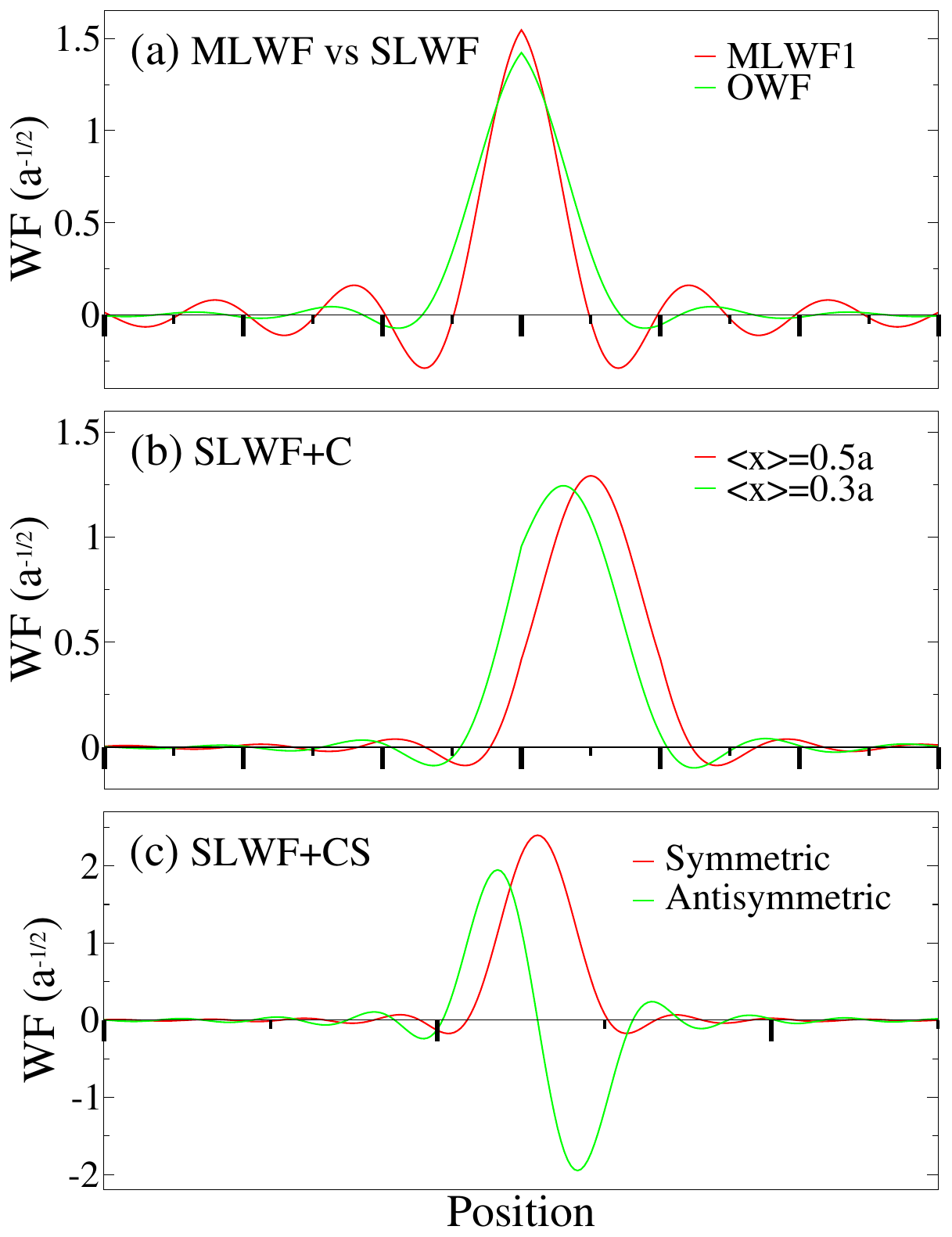}
\end{center}
\caption{
Wannier functions for the 1-d chain of negative $\delta$-function potentials. Large tick marks denote the $\delta$-function, while small tick marks denote the midpoint.
Panels (a) and (b) used $100$ $k$-points while panel (c) used $20$ $k$-points.
({\bf a}) Wannier functions obtained for $J=2,J'=2$ (MLWF) and $J=2,J'=1$ (SLWF). The spreads are $0.2713a^2$ and $0.0506a^2$, respectively.
({\bf b}) OWF with centers fixed at $0.5a$ and $0.3a$ (SLWF+C), in the case of $J=2,J'=1$. The spreads are $0.0596a^2$ and $0.0637a^2$, respectively.
({\bf c}) OWF with centers and symmetries controlled (SLWF+SC) in the case of $J=7, J'=1$. The spreads are $0.0050a^2$ and $0.0205a^2$ for symmetric OWF and antisymmetric OWF, respectively.
}
\label{negative-periodic}
\end{figure}

\section*{Appendix: One dimension with attractive delta potential \label{attractive delta}}

This Appendix presents results obtained for  a one dimensional chain of $\delta$ function potentials with negative values, i.e. the system considered in section \ref{positive-periodic} but with a change of sign in the potential. In this system a straightforward tight binding picture would be based on orbitals similar to the isolated delta-function bound states. We show that the SLWF procedure can be used to recover this picture, creating OWF that transforms according to the irreducible representations of the Hamiltonian. We also demonstrate that the SLWF procedure can be used to  generate states with symmetries not actually present in the Hamiltonian, provided that enough states are retained. 

In our analysis, we will consider two bands under a variety of different scenarios.  We begin by comparing MLWF ($J=2,J'=2$) with SLWF for the case of $J=2,J'=1$, and in Figure \ref{negative-periodic}(a) we plot the most localized MLWF and the objective Wannier function. In this case, both procedures naturally center the
Wannier functions at the potential and both orbitals transform like the identity.  As expected, the OWF has a smaller spread than the MLWF. In the second case, we perform SLWF+C for $J=2,J'=1$ (see Figure \ref{negative-periodic}(b)). First we center the OWF at the midpoint of the bond, successfully obtaining a symmetric function, though with a larger spread than the OWF which naturally centered
itself on the potential.  Subsequently, we chose to center the OWF  about a point $1/3$ of the way between the potentials, and this results in a similar spread and the Wannier function is  no longer symmetric about its center (see Figure \ref{negative-periodic}(b)).  If we perform SLWF+CS and attempt to enforce the OWF to be symmetric about its center, which is a symmetry that does not exist in the Hamiltonian, we were not successful (not shown). However, if we perform the same test using  $J=7,J'=1$ (see  Figure \ref{negative-periodic}(c)), there is much more freedom as we are only minimizing $1$ out of $7$ bands and a nearly symmetric function can be obtained.  Finally, we repeat the preceding case but demand an antisymmetric function, demonstrating that this is straightforward.

\end{document}